\documentclass[aps,prd,showpacs,superscriptaddress,nofootinbib,amsmath,amssymb,floats,floatfix,showkeys,notitlepage,longbibliography]{revtex4-1}

\usepackage{epsfig}
\usepackage{graphicx}
\usepackage{color}
\usepackage{enumitem}
\usepackage{amsmath}
\usepackage{amssymb}
\usepackage{adjustbox}
\usepackage{dcolumn}
\usepackage{bm}
\usepackage{hyperref}

\usepackage{multirow}
\usepackage{subcaption}
\usepackage{adjustbox}
\usepackage{float}
\begin{document}


\title{Quasinormal Modes and Bounding Greybody Factors of GUP--corrected Black Holes in Kalb--Ramond Gravity}

\author{Anshuman Baruah}\email{anshuman.baruah@aus.ac.in}
 \affiliation{Department of Physics, Assam University, Cachar - 788011, Assam, India}
 
\author{Ali \"Ovg\"un}%
 \email{ali.ovgun@emu.edu.tr}
\affiliation{%
 Physics Department, Eastern Mediterranean University, Famagusta, 99628 North Cyprus via Mersin 10, Turkey}%
\author{Atri Deshamukhya}\email{atri.deshamukhya@aus.ac.in}
 \affiliation{Department of Physics, Assam University, Cachar - 788011, Assam, India}

\date{\today}

\begin{abstract}
The vacuum expectation value of the non--minimally coupled Kalb--Ramond (KR) field leads to spontaneous local Lorentz symmetry violation, and static spherically symmetric solutions exist. In this study, we study the quasinormal modes (QNMs) of modified black holes in non--minimally coupled KR gravity. We employ a higher-order Pad\'e averaged WKB method to compute the QNMs for scalar, electromagnetic, and gravitational perturbations. In order to account for quantum corrections, we examine the geometric characteristics of the horizon and QNMs by introducing the generalized uncertainty principle (GUP). Additionally, we shed light on the impact of the Lorentz violating parameters on our findings and estimate QNMs for different perturbations. Further, we estimate bounds on the greybody factors for the modified and GUP--corrected black holes. Our findings reveal the influence of the Lorentz violating parameters in the model on the QNM frequencies and their reliance on the GUP parameters.
\end{abstract}

\keywords{Black holes; Quasinormal modes; Greybody factors; Lorentz symmetry violation; Kalb-Ramond gravity.}

\pacs{95.30.Sf, 04.70.-s, 97.60.Lf, 04.50.+h}

\maketitle


\section{Introduction}
\label{sec:int}
General Relativity (GR) is widely considered as one of the most extensively tested physical theories, and has been validated over the past century. Despite its success, it has certain limitations in that it cannot fully explain certain phenomena such as dark energy, and provides solely a classical depiction of gravitation. As a result, efforts to merge electromagnetism and GR can be traced back to the inception of GR \cite{Starobinsky:2007hu,Nojiri:2006ri,Copeland:2006wr,Sotiriou:2008rp,Clifton:2011jh,Bamba:2012cp,Nojiri:2010wj,Nojiri:2017ncd}. Moreover, in recent decades, various modifications and/or extensions to GR have been proposed to address its deficiencies. With recent advancements in experimental and observational technologies, it is now feasible to explore potential deviations from GR. While present observations of black holes from cutting--edge experiments, such as those conducted by the LIGO-Virgo and EHT collaborations, support GR predictions \cite{
LIGOScientific:2016aoc,LIGO16e,Event2021a,Event2021b}, it is anticipated that higher--sensitivity observations in the future may reveal deviations from GR. Thus, black holes offer an intriguing `laboratory' to study aspects of gravity in high--energy environments, and the study of observable quantities, such as gravitational wave signals and black hole shadows, from modified gravity is a subject of considerable interest \cite{Cardoso:2016rao,Cardoso:2016oxy,Cardoso:2017cqb,Vagnozzi:2022moj,Toshmatov18c,Pantig:2022gih,Lambiase:2023hng,Pantig:2022qak,Uniyal:2022vdu,Pantig:2022ely,Yang:2022ifo,Okyay:2021nnh,Gogoi:2023kjt,Ovgun:2023ego,Javed:2023iih,Kumaran:2022soh}.
\\
    
Modifications to Einstein's GR considering space--time torsion can be traced back to the work of Cartan \cite{cartan1, cartan2, cartan3, cartan4}. The Christoffel symbols (metric connections) are crucial in extracting meaningful information from curved space--times, and these are symmetric in the last two indices in GR \cite{wald2010general}. The fundamental difference from GR in Cartan's approach is the antisymmetry of the connections in the last two indices. Recent developments in String Theory, which is perhaps the best candidate of a quantum gravity theory, have indicated the presence of self--interacting tensor fields in the spectrum, which may spontaneously and locally break local Lorenz symmetry, in addition to inducing torsion in space--time. Unlike free fields governed by linear equations, self--interacting fields and potentials exhibit intriguing physical consequences. In the context of gravitation, one of the primary motivations for studying these modifications is that interactions may exhibit a preferred direction in space--time in the presence of these relic background fields, and future tests have the potential to reach sensitivities capable of measuring deviations from GR from distant galactic signals. In the context of heterotic string theory, one encounters two self--interacting tensor fields: the rank one Bumblebee field \cite{bumblebee, PhysRevD.69.105009} and the rank two Kalb--Ramond (KR) field \cite{KR}. The effective field theory of the non--minimally coupled KR field via the vacuum expectation value (VEV) was developed in Ref. \cite{KRVEVAction}, where the additional terms of interest in the action are the field strength, $H_{\mu \nu \lambda}$, and the KR field $B_{\mu \nu}$.
\\
    
A wide range of solutions have been reported in Bumblebee and KR gravity. For instance, static and non--static black hole solutions have been reported in Bumblebee gravity in Refs. \cite{Maluf:2013nva,Maluf:2014dpa,Maluf:2015hda,Maluf:2020kgf,Nascimento:2023auz,Capozziello:2023rfv,Khodadi:2023yiw,Xu:2023xqh,Filho:2022yrk,Xu:2022frb,Oliveira:2021abg,Poulis:2021nqh,Casana:2017jkc,Bluhm:2004ep,Delhom:2022xfo,Delhom:2020gfv,Delhom:2019wcm,Kuang:2022xjp,gullu2022schwarzschild, casana2018exact, jha2021bumblebee,Ovgun:2018xys,Li:2020dln,Khodadi:2022dff,Khodadi:2021owg}. The static solution in Ref. \cite{casana2018exact} context does not modify the Schwarzschild event horizon, which is in contrast to the model we will consider here, as demonstrated later. Gogoi et al. \cite{gogoi2022quasinormal} have presented comprehensive analyses of black hole quasinormal modes (QNMs) and properties of black holes in Bumblebee gravity. In the context of KR gravity, static, spherically symmetric solutions in the minimally coupled scenario have been reported, and it has been shown that these solutions exhibit features resembling Morris-Thorne wormholes \cite{Ovgun:2018xys,sengupta2001spherically, kar2003static}. Strong gravitational lensing and tests for extra dimensions with the KR field have been reported in Ref. \cite{chakraborty2017strong}. The non--minimal coupling of the KR VEV has been considered in Ref. \cite{lessa2020modified}, and Schwarzschild--like black hole solutions have been derived. Further, rotating black hole solutions have also been reported, and gravitational deflection of light and the shadow cast by rotating KR black holes have been studied in Ref.  \cite{kumar2020gravitational}. In this study, we aim to evaluate the QNMs of Schwarzschild--like black holes in non--minimally coupled KR gravity, and understand the effects of the Lorentz violating (LV) parameters on the results. Moreover, we estimate bounds on the greybody factors from these black holes.
\\

Black hole QNMs are solutions to the relevant perturbation equations of black hole perturbation theory, and describe ingoing waves at the horizon and outgoing waves at infinity \cite{Andersson:1992scr,Andersson:1994rm,Andersson:1995vi,Andersson:1996xw,Maggio:2019zyv,2022zym,Konoplya:2018yrp,Berti:2005ys}. The imaginary part of these complex frequencies yield the damping of the waves. QNMs are of particular interest in the context of gravitational wave (GW) astronomy, as the decaying tails of the GW signals can be observed with the current sensitives of the LIGO-Virgo experiments. The motivation for studying black hole QNMs in modified gravity is that as future detectors attain higher sensitivities, any deviations from GR Schwarzschild--like solutions may possibly be observed \cite{Daghigh:2008jz,Daghigh:2011ty,Daghigh:2020mog,Zhidenko:2003wq,Zhidenko:2005mv,Lepe:2004kv,Gonzalez:2017shu,Rincon:2021gwd,Gonzalez:2022ote,Panotopoulos:2020mii,Rincon:2020cos,Rincon:2018ktz,Fernando:2016ftj,Fernando:2015kaa,Fernando:2012yw,Fernando:2008hb,Khodadi:2021owg,Khodadi:2017eim,Khodadi:2018scn}. Further, the assumption of the existence of a minimum length scale demands corrections to black hole radii at energies reaching the Planck--scale. To this end, we incorporate quantum--gravity inspired generalized uncertainty principle (GUP) corrections to the studied black hole solution, and probe the differences from the original results, in addition to highlighting the effects of the GUP parameters \cite{Maggiore:1993rv,Kempf:1994su,Lambiase:2022xde,Ovgun:2015jna}.
\\

The remainder of the manuscript is organized as follows. In Section \ref{sec:two}, we review the KR field theory and derive the black hole solution to be studied. Further, we compute the horizon radii for different parametrizations, and introduce GUP correction to the black hole solution. Next, we introduce briefly the framework for evaluating the QNMs, and study the effects of different perturbations, viz. scalar, electromagnetic (EM), and gravitational perturbations in Sec. \ref{sec:three}. We obtain bounds on the greybody factors in Section 4. Finally, we present discussions and concluding remarks in Section \ref{sec:conc}.

\section{Static, spherically symmetric black hole solution in KR gravity}
\label{sec:two}
We consider the field theory proposed initially in Ref. \cite{KRVEVAction}, and following Lessa et al., \cite{lessa2020modified}, we briefly review the derivation of the Schwarzschild--like modified black hole solution. The KR field is an antisymmetric rank--2 tensor field arising in the bosonic string spectrum, and can be interpreted as a generalization of the electromagnetic potential with two indices. The 2--form potential and field strength are given as $B_2=\frac{1}{2} B_{\mu \nu} dx^{\mu} \land dx^\nu$ and $H_{\alpha\mu\nu}=\partial_{[\alpha}B_{\mu\nu]}$, respectively. As is well--known, $p$-form theories \cite{bandos2021p} are generalizations of Maxwell's electrodynamics, where the gauge potential is a higher--rank differential form \cite{wald2010general}. The Einstein-Hilbert action non--minimally
coupled with the self--interacting KR field is given by
\begin{equation}
S=\int \sqrt{-g}d^4x\left[\frac{R}{16\pi G}-\frac{1}{12}H_{\alpha\mu\nu}H^{\alpha\mu\nu}-V(B_{\mu\nu}B^{\mu\nu}
\pm b_{\mu\nu}b^{\mu\nu}) + \frac{1}{16\pi G}\left(\xi_2 B^{\mu\lambda}B^{\nu}_{\lambda}R_{\mu\nu}+\xi_3B_{\mu\nu}B^{\mu\nu}R\right) \right],
\label{eq:action}  
\end{equation}

where $\xi_i$ represent the non--minimal coupling constants. We consider the KR VEV with a constant norm and vanishing Hamiltonian. Varying Eq. \eqref{eq:action}, we obtain a modified form of the Einstein equations as
\begin{equation}
    G_{\mu \nu} = \kappa T_{\mu \nu}^{\xi_2}
    \label{eq:efe}
\end{equation}
with
\begin{eqnarray}
T_{\mu \nu}^{\xi_2} = \frac{\xi_2}{\kappa} \left[ \frac{1}{2} g_{\mu \nu} B^{\alpha \gamma}B^\beta{_\gamma} R_{\alpha \beta} - B^\alpha_\mu B^\beta_\nu R_{\alpha \beta} - B^{\alpha \beta}B_{\mu \beta} R_{\nu \alpha} - B^{\alpha \beta} B_{\nu \beta} R_{\mu \alpha} + \frac{1}{2} D_{\alpha}D_{\mu} (B_{\nu \beta} B^{\alpha \beta}) \right. \\ \nonumber + \left. \frac{1}{2} D_{\alpha}D_{\nu} (B_{\mu \beta} B^{\alpha \beta}) - \frac{1}{2} D^2 (B^\alpha_\mu B_{\alpha \nu}) - \frac{1}{2} g_{\mu \nu} D_\alpha D_\beta (B^{\alpha \gamma} B ^\beta_\gamma) \right]
\end{eqnarray}

A general static, spherically symmetric line element can be written as
\begin{equation}
    ds^2= -A(r) dt^2 + B(r) dr^2 + r^2 d\theta^2 + r^2 \sin ^2 \theta d\phi^2
    \label{eq:ssle}
\end{equation}

The previous assumption of the constancy of the KR VEV norm holds, provided that
\begin{equation}
    \Tilde{E}(r) = |b| \sqrt{\frac{A(r) B(r)}{2}},
\end{equation}

where $b$ is constant, and $E(r)$ defines a background pseudo--electric static field. Now, Eq. \eqref{eq:efe} can be recast as
\begin{align}
    R_{\mu \nu} = \xi_2 \left[g_{\mu \nu} b^{\alpha \gamma} b^\beta_\gamma R_{\alpha \beta} - b^\alpha_\mu b^\beta_\nu R_{\alpha \beta} - b^{\alpha \beta} b_{\mu \beta}R_{\nu \alpha} - b^{\alpha \beta} b_{\nu \beta} R_{\mu \alpha} + \frac{1}{2} D_\alpha D_\mu (b_{\nu \beta} b^{\alpha \beta}) \right. \\ \nonumber \left. + \frac{1}{2} D_\alpha D_\nu (b_{\mu \beta} b^{\alpha \beta} - \frac{1}{4} D^2(b^\alpha_\mu b_{\alpha \nu}))  \right]
\end{align}

Further, considering the metric \textit{ansatz} in Eq. \eqref{eq:ssle}, one can derive the following:
\begin{eqnarray}
\left(1-\frac{\lambda}{2} \right) R_{tt} = 0
\label{eq:Rtt}
\\
\left(1-\frac{\lambda}{2} \right) R_{rr} = 0
\label{eq:Rrr}
\\
R_{\theta \theta} = \frac{\lambda r^2}{2} \left( \frac{R_{tt}}{A(r)} - \frac{R_{rr}}{B(r)} \right)
\label{eq:Rthth}
\\
R_{\phi \phi} = sin^2\theta R_{\theta \theta}
\end{eqnarray}

Here, we use the redefinition $\lambda \equiv |b|^2 \xi_2$. Then, it is straight forward to match the corresponding components for the metric in Eq. \eqref{eq:ssle}, and using Eqs. \eqref{eq:Rtt} and \eqref{eq:Rrr} yields
\begin{equation}
    A(r)=\frac{1}{B(r)}
\end{equation}
for $\lambda \neq 2$, and using this as a constraint in Eq. \eqref{eq:Rthth}, we obtain
\begin{equation}
    \frac{r^2 \lambda}{2} A'' + (\lambda +1)r A' + A-1 =0
\end{equation}

The above equation is solved to obtain
\begin{equation}
    A(r) = 1- \frac{2GM}{r} + \frac{\Gamma}{r^{\frac{2}{\lambda}}}
\end{equation}

Thus, the power--law hairy (pseudo--electric) black hole metric takes the form \cite{lessa2020modified}
\begin{equation}
    ds^2 = - \left[1-\frac{R_s}{r} + \frac{\Gamma}{r^{\frac{2}{\lambda}}} \right]dt^2+\left[1-\frac{R_s}{r} + \frac{\Gamma}{r^{\frac{2}{\lambda}}} \right]^{-1} dr^2+r^2d\Omega^2
    \label{eq:BHle}
\end{equation}

In the above line element, the parameter $\Gamma$ appears as a constant of integration, and has dimensions $[\Gamma] = L^{\frac{2}{\lambda}}$. It is straight forward to see that the Schwarzschild solution is modified with the presence of these parameters. The Schwarzschild solution is recovered for a given $\Gamma$ when $\lambda \rightarrow 0$, where $|b|^2$ and $\xi_2 \rightarrow 0$. In the action given by Eq. \eqref{eq:action}, the self--interaction potential $V(B_{\mu\nu}B^{\mu\nu}
\pm b_{\mu\nu}b^{\mu\nu})$ triggers spontaneous symmetry breaking with a non--zero VEV $\langle B_{\mu \nu} \rangle =b_{\mu\nu}$ that breaks local Lorentz and diffeomorphim symmetry \cite{KRVEVAction}, and has a constant norm in the vacuum configuration \cite{KRVEVAction, lessa2020modified}. This allows one to define LV coefficients in terms of $b$ throughout the space--time \cite{KRVEVAction}. Thus, from the definition $\lambda \equiv |b|^2 \xi_2$, we refer to it as an LV parameter. Next, it is apparent from Eq. \eqref{eq:BHle} that the space--time resembles the Reissner--Nordstrom solution for $\lambda =1$. In this case, the pseudo--electric field is radial and constant with $E(r) = \frac{|b|}{\sqrt{2}}$, which is inconsistent considering a field from a localized charge \cite{lessa2020modified}. Thus, $\Gamma$ is interpreted not as a charge, but as an LV \emph{hair}.

\subsection{Horizons}
Horizons in spherically symmetric space--times can be identified as physically non--singular surfaces where $g_{tt} \rightarrow 0$. For the black hole solution Eq. \eqref{eq:BHle}, we identify horizons using the following master equation
\begin{eqnarray}
    \left(1-\frac{R_s}{r} + \frac{\Gamma}{r^{\frac{2}{\lambda}}} \right) = 0, \nonumber
    \\ 
    \Rightarrow r^{\frac{2}{\lambda}} - R_s r^{\frac{2}{\lambda}-1} + \Gamma =0,
\end{eqnarray}

which cannot be solved exactly for arbitrary $\lambda$. It can be verified that for $\lambda \rightarrow 0$ (the Lorentz invariant regime), we recover the Schwarzschild scenario with a solution $r=R_s$. Here, we solve for the horizon with different values of $\lambda$. For instance, with $\lambda = 1$, one obtains two solutions viz.
\begin{equation}
    r_{\pm} = \frac{R_s}{2} \left(1 \pm \sqrt{1-\frac{4\Gamma}{R_s^2}} \right)
    \label{eq:hor_one}
\end{equation}

\begin{figure}[htbp]
    \centering
\includegraphics[width=0.9\textwidth]{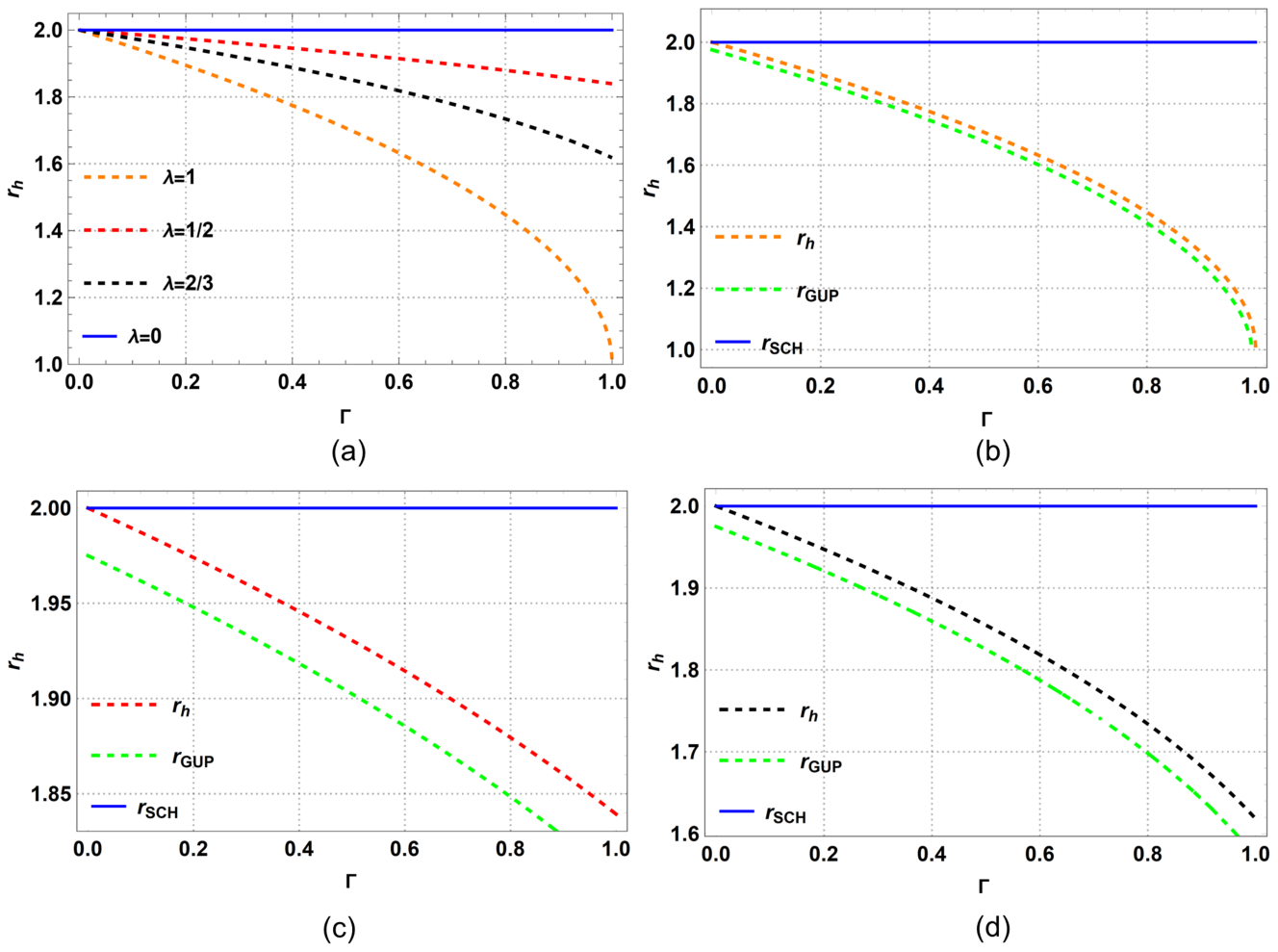}
    \caption{Profile of the black hole horizon. (a) With different choices of $\lambda$, (b) GUP--corrected horizon for $\lambda=1$, (c) GUP--corrected horizon for $\lambda=1/2$, (d) GUP--corrected horizon for $\lambda=2/3$. The $y-$axes are the horizon radii and $x-$axes are the LV parameter $\Gamma$. The blue line denotes the Schwarzschild horizon.}
    \label{fig:one}
\end{figure}

Here, we note some interesting properties of the black hole horizon. First, it is clear that the LV parameters modify the black hole horizon. Eq. \eqref{eq:hor_one} describes two solutions
\begin{eqnarray}
    r_+ \approx R_s - \frac{\Gamma}{R_s} - \frac{\Gamma^2}{R_s^3}
    \\
    r_- \approx \frac{\Gamma}{R_s} - \frac{\Gamma^2}{R_s^3}
\end{eqnarray}

considering $\Gamma << R_s^2$, where $r_-$ is a new inner horizon, and the outer (Schwarzschild) horizon is reduced by $R_s - r_-$. These new horizons are a unique feature, in contrast to the LSB Bumblebee black hole \cite{LSBBumBH}, where the Schwarzschild horizon is unchanged; and the Lorentz invariant KR black hole \cite{LorInvKRBH}, where the KR field changes the structure to a naked singularity. Further, we estimate the horizon radii with several values of $\lambda$ and check the variation with the parameter $\Gamma$ in Fig. \ref{fig:one} (a). It can be seen that the horizon radius approaches the Schwarzschild one as $\Gamma \rightarrow 0$. 

\subsection{GUP corrected solution}
Different well--discussed motivations exist that indicate the existence of a minimum measurable length scale in nature. For instance, a significant implication of gravitational physics is the thermodynamic interpretation of black holes \cite{bekenstein2020black, bekenstein1974generalized}, and subsequently, the prediction of black hole entropy and temperature \cite{hawking1974black, hawking1975particle}. Corrections to the black hole entropy at high--energy scales demand the introduction of correction coefficients that should preferably be estimated in a model--independent manner. Moreover, gravitational singularities, black hole scattering problems, and motivations arising from string theory indicate that the standard Heisenberg uncertainty principle should be modified in high--energy regimes. The conventional Heisenberg uncertainty principle does not take into account the effect of gravitation, which is consistent considering the weakness of gravity compared to the other fundamental forces. However, in strong--field regimes such as at the event horizons of black holes, it becomes necessary to consider gravitational effects. Considering quantum corrections to black hole event horizons motivates the introduction of the so--called generalized uncertainty principle (GUP), which can also be interpreted as modified commutation relations between position coordinates and momenta. The motivation for such modifications arises from the notion of introducing quantum corrections to black hole scattering problems, which is relevant considering the possibility of observing possible GW signatures arising from these effects. Several previous works have incorporated GUP corrections and analyzed properties of interest. There are several equivalent approaches of defining the GUP. For instance, one may write a GUP of the form \cite{PhysRevD.70.124021, ALI2009497, tawfik2014generalized, doi:10.1142/S0217751X1550030X, carr2015sub,Kempf:1994su, anacleto2020quantum, ali2009discreteness}
\begin{equation}
    \Delta x \Delta p \geq \frac{1}{2} \left(1 - \alpha \Delta p +\beta (\Delta p)^2  \right),
    \label{eq:gup1}
\end{equation}

where the Heisenberg uncertainty principle can be recovered for $\alpha = \beta = 0$ ($\alpha$ and $\beta$ are dimensionless and positive parameters). This inequality implies a minimum measurable length $\Delta x \geq (\Delta x)_{min} \approx (\sqrt{\beta} - \alpha)$ corresponding to a maximum measurable moment $\Delta p \leq (\Delta p)_{min} \approx \alpha/\beta$. Further, the Planck length corresponds to the smallest measurable length, and here, the position uncertainty approaches the Compton wavelength. Near the horizon, particles exhibit a wavelength of approximately the inverse of the Hawking temperature (in natural units), and thus, the position uncertainty is of the order of the horizon radius, $\Delta x \sim r_h$.
\\

Now, considering $\Delta p \sim p \sim E$, the Heisenberg inequality becomes $E \Delta x \geq \frac{1}{2}$. Using this, we can rearrange Eq. \eqref{eq:gup1} and consider a power law in $\alpha$ and $\beta$ to write:
\begin{equation}
    \mathcal{E} \geq E \left[1-\frac{\alpha}{(\Delta x)} + \frac{\beta}{2(\Delta x)^2} + ... \right]
\end{equation}

where $\mathcal{E}$ is identified as the GUP--corrected energy. Further, considering $ \Delta x \approx r_h$, we can write
\begin{equation}
    \mathcal{M} = M_{GUP} \geq M \left [1 - \frac{\alpha}{r_h} + \frac{\beta}{2 r_h^2} \right]
\end{equation}

where $M_{GUP}$ is the GUP--corrected black hole mass. Thus, the GUP--corrected horizon radius (for a Schwarzschild black hole) takes the form
\begin{equation}
    r_{GUP} = 2 M_{GUP} \geq r_h \left(1 - \frac{\alpha}{r_h} + \frac{\beta}{2 r_h^2} \right)
\end{equation}

In our case, $r_h$ is replaced by the corresponding solutions of the horizon equation Eq. \eqref{eq:hor_one}. For instance, with $\lambda =1$, the correction to the Schwarzschild horizon is $r_h = \frac{1}{2} (R_s + \sqrt{R_s^2 - 4\Gamma})$, and the GUP--corrected horizon in terms of the corrected mass would be at $r_{GUP} = M_{GUP} + \sqrt{M_{GUP}^2-\Gamma}$. The GUP--corrected black hole metric can then be defined by replacing the mass in Eq. \eqref{eq:BHle} with $M_{GUP}$ as:

\begin{equation}
    ds^2 = - \left[1-\frac{2M_{GUP}}{r} + \frac{\Gamma}{r^{\frac{2}{\lambda}}} \right]dt^2+\left[1-\frac{2 M_{GUP}}{r} + \frac{\Gamma}{r^{\frac{2}{\lambda}}} \right]^{-1} dr^2+r^2d\Omega^2
    \label{GUPmetric}
\end{equation}

In Figs. \ref{fig:one} (b)-(d), we show comparisons of the horizon radii vs. the GUP--corrected radii for different values of $\Gamma$. In all cases, we see that the GUP--corrected horizon is smaller than the horizon of the modified KR black hole. Moreover, the horizons tend to the Schwarzschild radius as $\Gamma \rightarrow 0$, as expected. Unlike Schwarzschild black holes where the horizon radius is effectively increased due to GUP-corrections, our results show that the horizon radius decreases upon introducing GUP correction.
\section{Quasinormal modes}
\label{sec:three}
Recent advances in GW astronomy have ushered in a new era of multi-messenger astronomy. Anticipating future advancements in this field, it is expected that any deviations from GR may potentially be tested in the near future with the development of high--sensitivity GW detectors. To this end, it becomes crucial to estimate GW signatures from novel models that may serve as references for comparing future results with theory. Especially, gravitational perturbations are closely related to the ringdown signals emitted. Here, we report for the first time the QNMs of modified black hole solutions in non--minimally coupled KR gravity. QNMs are solutions of relevant black hole perturbation equations, which can be studied considering perturbations of test fields\footnote{The back-reaction of the fields on the space--time is considered as negligible} such as scalar and vector (electromagnetic) fields, and also considering gravitational perturbations. The governing equations are encoded in the corresponding equations of motion of the fields considered. For scalar perturbations, one can start with the Klein--Gordon equation, and for EM and gravitational perturbations, the tetrad formalism may be leveraged. Considering axial perturbations, the perturbed metric can be recast without loss of generality as follows \cite{bouhmadi2020consistent}:

\begin{equation}
    ds^2 = -|g_{tt}| dt^2 + g_{rr}dr^2 + r^2 d\theta^2 + r^2\sin^2 \theta \left(d\phi - a dt - bdr - c d\theta \right)^2
    \label{pertmetric}
\end{equation}

Here, $a$, $b$, and $c$ are functions of $t$, $r$, and $\theta$, encoding the perturbations. We leverage the tetrad formalism and adopt a basis $e^\mu_{a}$ associated with the metric $g_{\mu\nu}$, and satisfying
\begin{align}
e^{(a)}_\mu e^\mu_{(b)} &= \delta^{(a)}_{(b)} \notag \\
e^{(a)}_\mu e^\nu_{(a)} &= \delta^{\nu}_{\mu} \notag \\
e^{(a)}_\mu &= g_{\mu\nu} \eta^{(a)(b)} e^\nu_{(b)}\notag \\
g_{\mu\nu} &= \eta_{(a)(b)}e^{(a)}_\mu e^{(b)}_\nu = e_{(a)\mu} e^{(a)}_\nu.
\end{align}

In the new basis, vector and tensor quantities are projected as
\begin{align}
P_\mu &= e^{(a)}_\mu P_{(a)}, \notag\\ 
P_{(a)} &= e^\mu_{(a)} P_\mu, \notag\\
A_{\mu\nu} &=  e^{(a)}_\mu e^{(b)}_\nu A_{(a)(b)}, \notag\\
A_{(a)(b)} &= e^\mu_{(a)} e^\nu_{(b)} A_{\mu\nu}.
\end{align}

The implementation of the tetrad formalism \cite{chandrasekhar1991selected} in this perturbed space--time is discussed in the following subsections.

\subsection{Massless Scalar Perturbations}
Considering the propagation of a massless scalar field around the black hole and assuming that the reaction of 
the scalar field on the space--time is negligible, the scalar QNMs can be described by the Klein--Gordon equation given by
\begin{equation}\label{scalar_KG}
\square \Phi = \dfrac{1}{\sqrt{-g}} \partial_\mu (\sqrt{-g} g^{\mu\nu} \partial_\nu \Phi) = 0.
\end{equation}

Neglecting the back--reaction of the field, one can consider Eq. \eqref{pertmetric} only up to the zeroth order:
\begin{equation}
    ds^2 = -|g_{tt}| dt^2 + g_{rr}dr^2 + r^2 d\Omega_2^2
\end{equation}

The scalar field can conventionally be decomposed using spherical harmonics as
\begin{equation}
\Phi(t,r,\theta, \phi) = \dfrac{1}{r} \sum_{l,m} \psi_l(t,r) Y_{lm}(\theta, \phi),
\end{equation}

where $\psi_l(t,r)$ is the time--dependent radial wave function. Further, $l$ and $m$ are 
the indices of the spherical harmonics $Y_{lm}$. Then, Eq. \eqref{scalar_KG} yields
\begin{equation}
\partial^2_{r_*} \psi(r_*)_l + \omega^2 \psi(r_*)_l = V_s(r) \psi(r_*)_l,  
\end{equation}

where $r_*$ is the tortoise coordinate defined as
\begin{equation}\label{tortoise}
\dfrac{dr_*}{dr} = \sqrt{g_{rr}\, |g_{tt}^{-1}|}
\end{equation}

and $V_s(r)$ is the effective potential of the field given by
\begin{equation}\label{Vs}
V_s(r) = |g_{tt}| \left( \dfrac{l(l+1)}{r^2} +\dfrac{1}{r \sqrt{|g_{tt}| g_{rr}}} \dfrac{d}{dr}\sqrt{|g_{tt}| g_{rr}^{-1}} \right).
\end{equation}

\subsection{Electromagnetic Perturbations}
Considering EM perturbations in the tetrad formalism \cite{chandrasekhar1991selected}, one can write from the Bianchi identity of the EM field strength $F_{[(a)(b)(c)]} = 0$:
\begin{align}
\left( r \sqrt{|g_{tt}|}\, F_{(t)(\phi)}\right)_{,r} + r \sqrt{g_{rr}}\, F_{(\phi)(r), t} &=0, \label{em1} \\
\left( r \sqrt{|g_{tt}|}\, F_{(t)(\phi)}\sin\theta\right)_{,\theta} + r^2 \sin\theta\, F_{(\phi)(r), t} &=0. \label{em2}
\end{align}

Additionally, we have the conservation equation, $\eta^{(b)(c)}\! \left( F_{(a)(b)} \right)_{|(c)} =0$, which gives
\begin{equation} \label{em3}
\left( r \sqrt{|g_{tt}|}\, F_{(\phi)(r)}\right)_{,r} +  \sqrt{|g_{tt}| g_{rr}}\, F_{(\phi)(\theta),\theta} + r \sqrt{g_{rr}}\, F_{(t)(\phi), t} = 0.
\end{equation}

Redefining the field perturbation as $\mathcal{F} = F_{(t)(\phi)} \sin\theta$, we can differentiate Eq. \eqref{em3} w.r.t.\ $t$ and use equations \eqref{em1} and \eqref{em2} to get
\begin{equation}\label{em4}
\left[ \sqrt{|g_{tt}| g_{rr}^{-1}} \left( r \sqrt{|g_{tt}|}\, \mathcal{F} \right)_{,r} \right]_{,r} + \dfrac{|g_{tt}| \sqrt{g_{rr}}}{r} \left( \dfrac{\mathcal{F}_{,\theta}}{\sin\theta} \right)_{,\theta}\!\! \sin\theta - r \sqrt{g_{rr}}\, \mathcal{F}_{,tt} = 0,
\end{equation}

Then, using the Fourier decomposition $(\partial_t \rightarrow -\, i \omega)$ and field decomposition\footnote{Here, $Y(\theta)$ is the Gegenbauer function \cite{abramowitz1964handbook}.} $\mathcal{F}(r,\theta) = \mathcal{F}(r) Y_{,\theta}/\sin\theta$ \cite{chandrasekhar1991selected}, Eq. \eqref{em4} can be recast as
\begin{equation}\label{em5}
\left[ \sqrt{|g_{tt}| g_{rr}^{-1}} \left( r \sqrt{|g_{tt}|}\, \mathcal{F} \right)_{,r} \right]_{,r} + \omega^2 r \sqrt{g_{rr}}\, \mathcal{F} - |g_{tt}| \sqrt{g_{rr}} r^{-1} l(l+1)\, \mathcal{F} = 0.
\end{equation}

Finally, using $\psi_e \equiv r \sqrt{|g_{tt}|}\, \mathcal{F}$ and introducing the tortoise coordinate defined previously, the perturbation equation can be described by a Schr\"odinger-like equation given by
\begin{equation}
\partial^2_{r_*} \psi_e + \omega^2 \psi_e = V_e(r) \psi_e,
\end{equation}
with the potential defined as
\begin{equation}\label{Ve}
V_e(r) = |g_{tt}|\, \dfrac{l(l+1)}{r^2}. 
\end{equation}

\subsection{Gravitational Perturbations}
To evaluate axial gravitational perturbations in effective theories such as the one in our framework, it can be considered that the black hole is described by Einstein gravity minimally coupled to an anisotropic source, and perturbations can be considered as encoded in the perturbations of the gravitational field equation and the corresponding anisotropic energy--momentum tensor. In the tetrad formalism, the axial components of perturbed energy--momentum tensor are zero \cite{Chen:2019iuo}, and thus, the master equation can be derived from $R_{(a)(b)} = 0$; the $\theta \, \phi$ and $r\, \phi$ components yield \cite{bouhmadi2020consistent}
\begin{align}
\left[ r^2 \sqrt{|g_{tt}| g_{rr}^{-1}}\, (b_{,\theta} - c_{,r}) \right]_{,r} = r^2 \sqrt{|g_{tt}|^{-1} g_{rr}}\, (a_{,\theta} - c_{,t})_{,t}, \label{g1}\\
\left[ r^2 \sqrt{|g_{tt}| g_{rr}^{-1}}\, (c_{,r} - b_{,\theta}) \sin^3\theta  \right]_{,\theta} = \dfrac{r^4 \sin^3\theta}{\sqrt{|g_{tt}| g_{rr}}}\, (a_{,r} - b_{,t})_{,t}. \label{g2}
\end{align}

Now, considering the ansatz $\mathcal{F}_g (r, \theta) = \mathcal{F}_g (r) Y(\theta)$, using the redefinition $\psi_g r = \mathcal{F}_g$, and introducing the tortise coordinate defined earlier, the master perturbation equation can be derived from Eqs. \eqref{g1} and \eqref{g2} as
\begin{equation}
\partial^2_{r_*} \psi_g + \omega^2 \psi_g = V_g(r) \psi_g,
\end{equation}

where the effective potential is given by
\begin{equation}\label{Vg}
V_g(r) = |g_{tt}| \left[ \dfrac{2}{r^2} \left( \dfrac{1}{g_{rr}} - 1 \right) + \dfrac{l(l+1)}{r^2} - \dfrac{1}{r \sqrt{|g_{tt}| g_{rr}}} \left( \dfrac{d}{dr} \sqrt{|g_{tt}| g_{rr}^{-1}} \right) \right].
\end{equation}

\begin{figure}[htbp]
    \centering
    \includegraphics[width=0.6\textwidth]{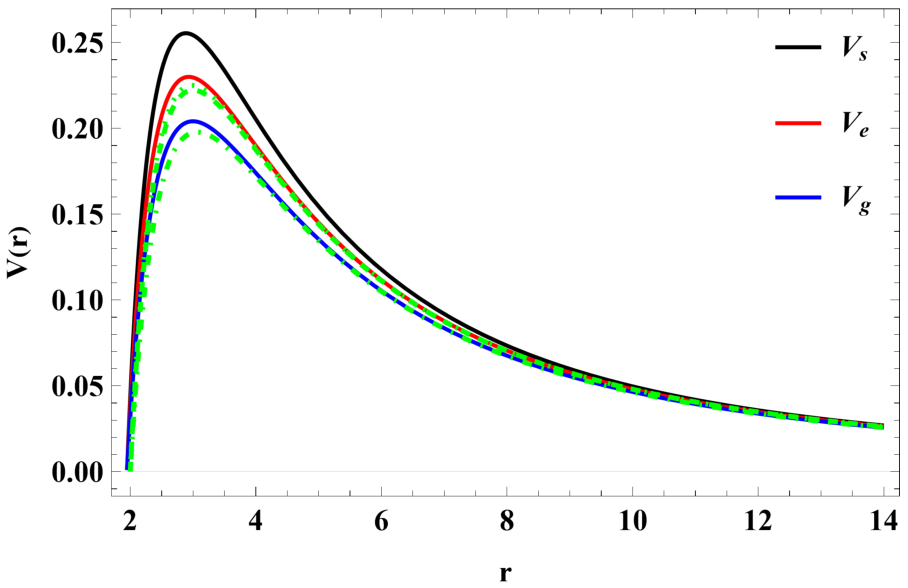}
    \caption{Effective potentials for the scalar (black), electromagnetic (red), and gravitational (blue) cases with $M=1$, $l=2$, $\Gamma =0.1$, and $\lambda =1$. The Schwarzschild behavior ($\lambda = \Gamma=0$) is shown by the green dotted, dashed, and dot--dashed lines for scalar, electromagnetic, and gravitational perturbations, respectively.}
    \label{fig:poten}
\end{figure}

The profiles of $V_s$, $V_e$, and $V_g$ have been shown in Fig. \ref{fig:poten}.

\subsection{Pad\'{e}-averaged WKB method for QNMs}
\label{sec:padeqnm}
To solve the perturbation equations in the frequency domain, the WKB approach is the most widely discussed method. The WKB method for black hole perturbations was developed to first order by Mashoon \cite{mashoon} and Schutz and Will \cite{schutz1985black}, to third order by Iyer and Will \cite{iyer1987black}, up to the sixth order by Konoplya \cite{Konoplya6thOrder}, and up to the 13\textsuperscript{th} order in \cite{matyjasekopalaWKB}. However, it is well--known that the WKB approach yields inaccurate results for $n \geq l$, as demonstrated below. The Pad\'{e} approximation \cite{matyjasekopalaWKB} has recently been used to evaluate QNMs with higher precision. Here, we employ the Pad\'{e}-averaged WKB approach to estimate the QNMs and compare the results by estimating the relative error. A brief overview of the WKB method as outlined in Refs. \cite{Konoplya6thOrder} and \cite{konoplya2019higher} has been presented below.\footnote{We refer the readers to Ref. \cite{konoplya2011quasinormal} for a detailed review of the WKB method for black hole perturbations, and to the original works in Refs. \cite{Konoplya6thOrder, konoplya2019higher} for further details on the method employed here.}
\begin{figure}[htbp]
    \centering
    \includegraphics[width=0.9\textwidth]{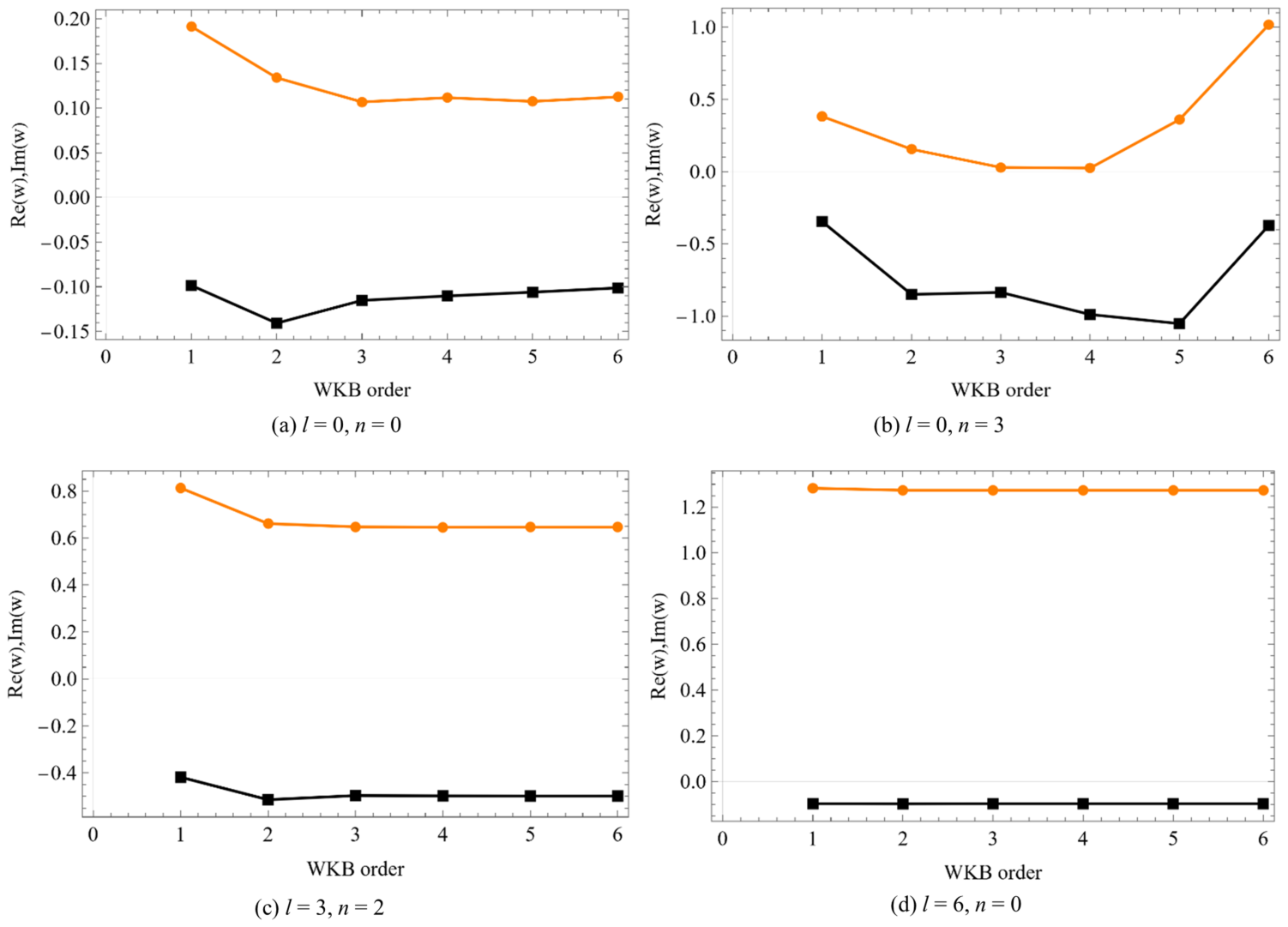}
    \caption{A demonstration of the convergence of the WKB formula for scalar perturbations with $\lambda = 1$ and $\Gamma = 0.1$. (a) $l=0,n=0$ \textrightarrow values fluctuate, (b) $l=0,n=3$ \textrightarrow values fluctuate significantly (c) $l=3,n=3$ \textrightarrow values converge approximately at higher orders (d) $l=6,n=0$ \textrightarrow values converge after 3\textsuperscript{rd} WKB order}.
    \label{fig:two}
\end{figure}

Considering a wave--like equation
\begin{equation}
    \label{eq:wave}
    \frac{d^2 \Psi}{dx^2} = U (x, \omega) \Psi,
\end{equation}

the WKB method is used to obtain solutions in the asymptotic regions described by a superposition of ingoing and outgoing waves \cite{konoplya2011quasinormal}. The effective potential $U (x, \omega)$ is asymptotically constant, has a single peak, with negative asymptotics. The WKB formula matches the asymptotic solutions to this equation with the extremum of the effective potential via a Taylor expansion. The general WKB formula in this case can be given by \cite{konoplya2011quasinormal} 
\begin{equation}
    \label{eq:genwkb}
    0=U_0(\omega)+A_2( \mathcal{K}^2 )+A_4(\mathcal{K}^2)+A_6( \mathcal{K}^2 )+\ldots- i \mathcal{K}\sqrt{-2U_2(\omega)}(1+A_3(\mathcal{K}^2)+A_5( \mathcal{K}^2 )+A_7( \mathcal{K}^2 )\ldots),
\end{equation}
where $A_k(\mathcal{K}^2)$ denote the corrections of order $k$ to the eikonal formula. The QNMs can then be obtained by analytic continuation of the above to the complex plane. In the case of black hole perturbations, Eq. \ref{eq:wave} is defined it terms of the tortoise coordinate, as described in the previous section. For spherically symmetric black holes, the general dependence of the effective potential on the frequency is of the form $U(x, \omega) = V(x) - \omega^2$, and perturbations of gravitational and test fields can be interpreted as superposition of multipoles on the $2-$sphere of the line--element given by Eq. \eqref{eq:ssle} \cite{konoplya2011quasinormal}. Finally, Eq. \eqref{eq:genwkb} provides a closed--form formula for the QNM frequencies as \cite{konoplya2019higher}
\begin{equation}
    \label{eq:bhwkb}
    \omega^2=V_0+A_2(\mathcal{K}^2)+A_4(\mathcal{K}^2)+A_6(\mathcal{K}^2)+\ldots- i  \mathcal{K}\sqrt{-2V_2}\left(1+A_3(\mathcal{K}^2)+A_5(\mathcal{K}^2)+A_7(\mathcal{K}^2)\ldots\right),
\end{equation}
where $\mathcal{K}$ admits half--integer values
\begin{eqnarray}
\mathcal{K} &=& \left\{
\begin{array}{ll}
 +n+\frac{1}{2}, & Re(\omega)>0; \\
 -n-\frac{1}{2}, & Re(\omega)<0; \phantom{\frac{{}^{Whitespace}}{}}
\end{array}
\right.\\\nonumber
&&\qquad\quad\qquad n=0,1,2,3\ldots.
\end{eqnarray}
Now, Pad\'{e} approximants can be used to improve the accuracy of this formula \cite{matyjasekopalaWKB}, where a polynomial $P_k(\epsilon)$ is defined in powers of the \emph{order parameter} $\epsilon$, modifying Eq. \eqref{eq:bhwkb} as
\begin{equation}
P_k(\epsilon)=V_0+A_2(\mathcal{K}^2)\epsilon^2+A_4(\mathcal{K}^2)\epsilon^4+A_6(\mathcal{K}^2)\epsilon^6+\ldots - i \mathcal{K}\sqrt{-2V_2}\left(\epsilon+A_3(\mathcal{K}^2)\epsilon^3+A_5(\mathcal{K}^2)\epsilon^5\ldots\right)
\end{equation}
Here, the polynomial order $k$ is the same as that for the WKB formula, and the squared frequency can be obtained with $\epsilon =1$ via $\omega^2=P_{\tilde{n}/\tilde{m}}(1)$, where $P_{\tilde{n}/\tilde{m}}(\epsilon)$ are Pad\'{e} approximants for the polynomial $P_k(\epsilon)$, given by rational functions of form \cite{matyjasekopalaWKB, konoplya2019higher}
\begin{equation}
    P_{\tilde{n}/\tilde{m}}(\epsilon)=\frac{Q_0+Q_1\epsilon+\ldots+Q_{\tilde{n}}\epsilon^{\tilde{n}}}{R_0+R_1\epsilon+\ldots+R_{\tilde{m}}\epsilon^{\tilde{m}}},
\end{equation}
with $\tilde{n}+\tilde{m}=k$. We employ this Pad\'{e} averaged WKB approach to estimate the QNMs, and the results have been presented below. The numerical approach following Ref. \cite{konoplya2019higher} can be implemented via the code found in \cite{konoplyacode}.
\\

Figure \ref{fig:two} shows an example of QNMs calculated using the WKB formula for scalar perturbations with $\lambda = 1$ and $\Gamma = 0.1$. It can be seen from the figure that the WKB formula performs worse for scalar perturbations when $n \geq l$, and the values converge for $l > n$. Thus, higher orders do not necessarily guarantee higher accuracy for the WKB approximation, and this is where the advantage of the Pad\'{e}-averaged WKB formula comes in handy. Hereafter, we focus on elucidating the effects of the LV parameters on the different QNMs. To this end, we compute and compare the QNMs for different perturbations. To estimate the accuracy of the approach, we estimate the errors in the calculation. In the WKB formula developed in Ref. \cite{Konoplya6thOrder}, corrections in each order affect the real and imaginary parts of the squared frequency. Thus, the error in $\omega_k$ for an arbitrary order $k$ can be estimated as
\begin{equation}
    \Delta_k = \frac{|\omega_{k+1} - \omega_{k-1}|}{2}
\end{equation}

Additionally, the table below shows how the WKB approximation worsens for higher orders.
\begin{ruledtabular}
\begin{table}[htbp]
    \centering
    \begin{tabular}{ccc}

\text{WKB order} & \text{quasinormal frequency} & \text{error estimation} \\
\hline
 12 & $28.3995\, -0.172994 i$ & 71.7859 \\
 11 & $2.23923\, -2.19404 i$ & 13.97 \\
 10 & $0.459599\, -0.104455 i$ & 1.36452 \\
 9 & $0.4921\, -0.0975565 i$ & 0.0166149 \\
 8 & $0.492042\, -0.097268 i$ & 0.000147062 \\
 7 & $0.492047\, -0.0972671 i$ & $4.6412\times 10^{-6}$ \\
 6 & $0.492049\, -0.097275 i$ & $8.6085\times 10^{-6}$ \\
 5 & $0.492063\, -0.0972721 i$ & 0.0000248217 \\
 4 & $0.492054\, -0.0972257$ i & 0.000218501 \\
 3 & $0.491628\, -0.09731 i$ & 0.00188962 \\

    \end{tabular}
    \caption{Variation of $l=2, n=0$ scalar QNMs with WKB order}
    \label{tab:one}
\end{table}
\end{ruledtabular}
\begin{table}[htbp]
    \begin{subtable}[t]{0.45\textwidth}
        \centering
             \adjustbox{width=\textwidth}{\begin{tabular}{|ccc|}

                \hline
                \text{order} & \text{frequency} & \text{error estimation} \\
                \hline
                1 & $0.497022\, -0.0933453 i$ & 0.0178376 \\
                2 & $0.492058\, -0.0968007 i$ & 0.00115709 \\
                3 & $0.49198\, -0.0972148 i$ & 0.0000190299 \\
                4 & $0.492055\, -0.0972625 i$ & $9.43967\times10^{-6}$ \\
                5 & $0.492055\, -0.0972708 i$ & $4.21641 \times10^{-6}$ \\
                6 & $0.49205\, -0.0972682 i$ & $5.76124 \times 10^{-7}$ \\
                7 & $0.49205\, -0.0972678 i$ & $7.54884 \times10^{-8}$ \\
                8 & $0.49205\, -0.0972682 i$ & $6.8456 \times 10^{-7}$ \\
                9 & $0.49205\, -0.0972679 i$ & $3.58757\times10^{-8}$ \\
                10 & $0.492048\, -0.0972686 i$ & $3.94756\times10^{-6}$ \\
                11 & $0.49205\, -0.097268 i$ & $2.87951\times10^{-7}$ \\
                12 & $0.492049\, -0.0972679 i$ & $1.75108\times10^{-6}$ \\
                13 & $0.49205\, -0.0972679 i$ & $3.02785\times10^{-7}$ \\
                \hline
                \end{tabular}
                }
                \caption{Variation of $l=2, n=0$ scalar QNMs with the Pad\'{e}-averaged WKB method}
                \label{tab:scalarpade}
                \end{subtable}
    \begin{subtable}[t]{0.45\textwidth}
        \centering
         \adjustbox{width=\textwidth}{
      \begin{tabular}{|ccc|}
\hline
 \text{order} & \text{frequency} & \text{error estimation} \\
 \hline
 1 & $0.471105\, -0.0914567 i$ & 0.0180862 \\
 2 & $0.465899\, -0.0950303 i$ & 0.00120512 \\
 3 & $0.465796\, -0.0954905 i$ & 0.0000187211 \\
 4 & $0.465873\, -0.0955419 i$ & $9.7659\times10^{-6}$ \\
 5 & $0.465871\, -0.0955505 i$ & $4.0325\times10^{-6}$ \\
 6 & $0.465867\, -0.0955477 i$ & $5.9339\times10^{-7}$ \\
 7 & $0.465867\, -0.0955474 i$ & $2.6084\times10^{-8}$ \\
 8 & $0.465868\, -0.0955491 i$ & $1.19426\times10^{-6}$ \\
 9 & $0.465867\, -0.0955474 i$ & $2.4867\times10^{-8}$ \\
 10 & $0.465865\, -0.0955482 i$ & $3.7704\times10^{-6}$ \\
 11 & $0.465867\, -0.0955474 i$ & $9.4381\times10^{-8}$ \\
 12 & $0.465867\, -0.0955489 i$ & $2.2377\times^{-6}$ \\
 13 & $0.465868\, -0.0955475 i$ & $4.2845\times10^{-7}$ \\
\hline
    \end{tabular}
    }
    \caption{Variation of EM QNMs with Pad\'{e} averaged WKB order with $l=2, n=0$}
    \label{tab:three}  
     \end{subtable}

     \begin{subtable}[t]{0.45\textwidth}
            \centering
         \adjustbox{width=\textwidth}{

    \begin{tabular}{|ccc|}
\hline
 \text{order} & \text{frequency} & \text{error estimation} \\
 \hline
 1 & $0.443262\, -0.0892795 i$ & 0.0183433 \\
 2 & $0.437784\, -0.0929841 i$ & 0.00126026 \\
 3 & $0.437645\, -0.0935049 i$ & 0.0000178982 \\
 4 & $0.43772\, -0.0935581 i$ & $9.42377\times10^{-6}$ \\
 5 & $0.437716\, -0.0935635 i$ & $2.37098\times10^{-7}$ \\
 6 & $0.437716\, -0.0935632 i$ & $6.20917\times10^{-8}$ \\
 7 & $0.437716\, -0.0935631 i$ & $5.95347\times10^{-8}$ \\
 8 & $0.437716\, -0.0935632 i$ & $6.21554\times10^{-8}$ \\
 9 & $0.437715\, -0.0935635 i$ & $1.57578\times10^{-7}$ \\
 10 & $0.437715\, -0.093564 i$ & $1.46729\times^10{-6}$ \\
 11 & $0.437716\, -0.0935631 i$ & $3.79497\times10^{-7}$ \\
 12 & $0.437715\, -0.093564 i$ & $1.43755\times10^{-6}$ \\
 13 & $0.437716\, -0.0935634 i$ & $7.19913\times10^{-7}$ \\
 \hline

    \end{tabular}
    
    }
    \caption{Gravitational QNMs using the Pad\'{e}-averaged WKB method with $l=2, n=0$ and $\Gamma = 0.1, \lambda =1$}
    \label{tab:five}
       \end{subtable}
       \begin{subtable}[t]{0.45\textwidth}
            \centering
         \adjustbox{width=\textwidth}{

    \begin{tabular}{|ccc|}
\hline
 \text{$l$} & \text{frequency} & \text{error estimation} \\
 \hline
  \multicolumn{3}{|c|}{Scalar}\\
   \cline{1-3}
    3 & $0.687078\, -0.0970139 i$ & $1.8665\times10^{-7}$ \\
    4 & $0.882443\, -0.0969075 i$ & $2.7904\times10^{-8}$ \\
    5 & $1.07796\, -0.0968536 i$ & $1.71102\times10^{-9}$ \\
    6 & $1.27356\, -0.0968226 i$ & $6.25243\times10^{-10}$ \\
 \hline\hline
  \multicolumn{3}{|c|}{EM}\\
   \hline
    3 & $0.668513\, -0.0961463 i$ & $4.18055\times10^{-8}$ \\
    4 & $0.868046\, -0.0963854 i$ & $3.43649\times10^{-8}$ \\
    5 & $1.0662\, -0.0965049 i$ & $1.81895\times10^{-9}$ \\
    6 & $1.26361\, -0.0965732 i$ & $5.62649\times10^{-10}$ \\
\hline\hline
  \multicolumn{3}{|c|}{Gravitational}\\
   \hline
    3 & $0.648926\, -0.0951982 i$ & $3.71152\times10^{-8}$ \\
    4 & $0.852971\, -0.0958268 i$ & $7.24278\times10^{-9}$ \\
    5 & $1.05393\, -0.0961359 i$ & $1.9821\times10^{-9}$ \\
    6 & $1.25326\, -0.096311 i$ & $7.65833\times10^{-10}$ \\
 \hline

    \end{tabular}
    
    }
    \caption{$n=0$ QNMs with varying $l$ and $\Gamma = 0.1, \lambda =1$}
    \label{tab:ss}
       \end{subtable}
     \caption{$n=0$ QNMs with the Pad\'{e}-averaged WKB method}
     \label{tab:QNMPadeModBH}
\end{table}

Table \ref{tab:QNMPadeModBH} shows the calculated $n=0$ QNM frequencies using Pad\'{e} approximation for scalar, EM, and gravitational perturbations and the associated errors in the frequencies. In subtable (d), we also show the frequencies calculated for different values of the multipole number. It can be seen from the data that we have good accuracy of estimation of the QNMs at higher orders. Overall, the estimated errors also decrease as we move to higher orders, indicating that the accuracy of the calculations improves with additional terms in the expansion. The behavior for scalar, EM, and gravitational perturbation are nearly identical. Precisely, $Re(\omega)$ follows the trend scalar$>$EM$>$gravitational; and Im$(\omega)$ follows gravitational$>$EM$>$scalar, indicating that gravitational perturbations are slow decaying. Comparing with the data for the $\lambda = \Gamma = 0$ (Schwarzschild) case (see Appendix \ref{appenA}), it can be seen that the QNMs for the modified black hole are approximately identical to those of the Schwarzschild black hole, and thus, any deviations may be difficult to verify. This trend can also be verified from the data in subtable \textbf{(d)}, where we again have good error bounds at higher multipole numbers for the $n=0$ overtone. Additionally, we show the frequency data at varying overtones in Table \ref{tab:newthree}. Next, we wish to check what are the effects of the LV parameters on the QNM frequencies.

\subsubsection{Variation of QNMs with $\Gamma$ and $\lambda$}
\begin{figure}[!h]
    \centering
    \includegraphics[width=0.9\textwidth]{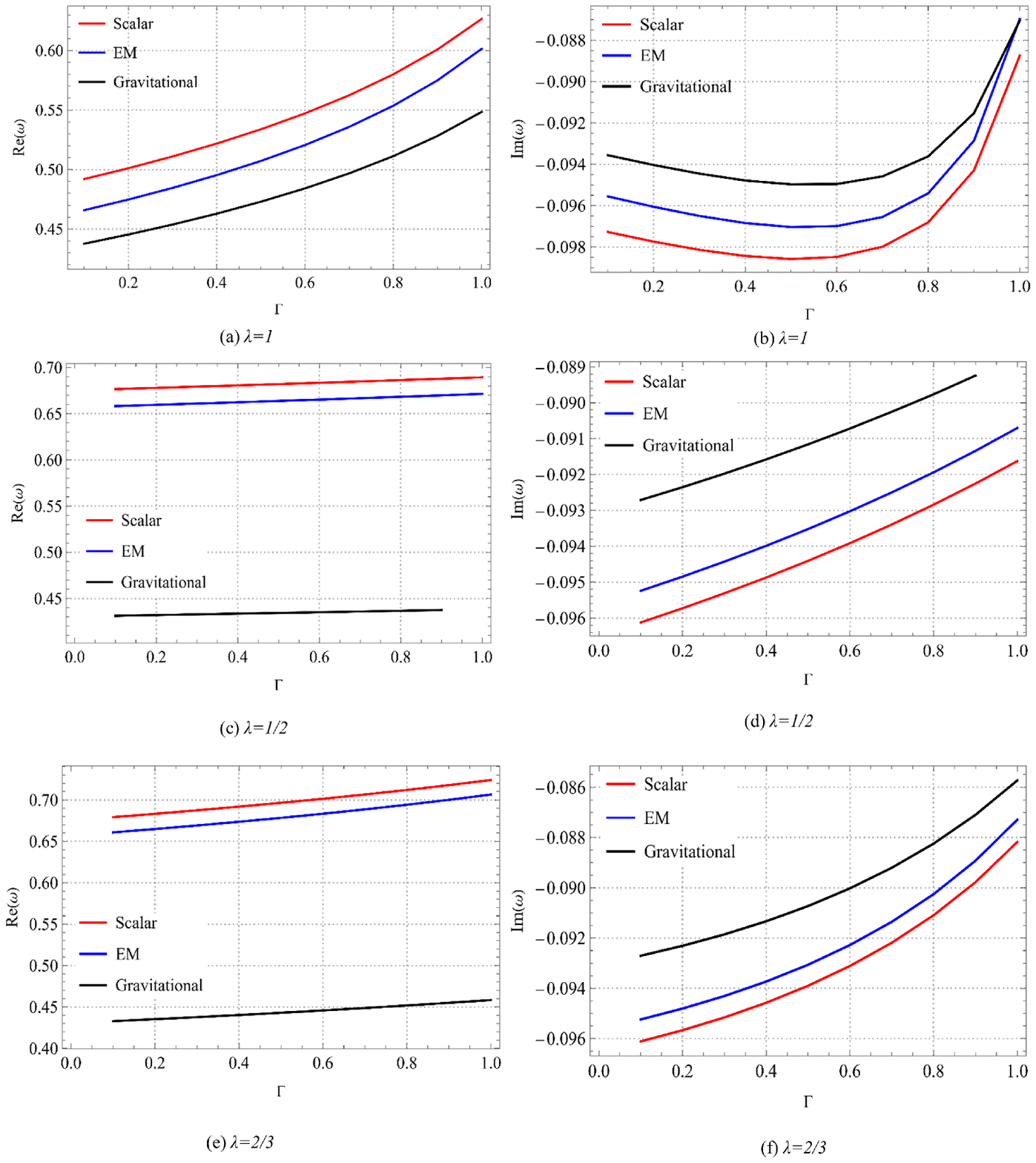}
    \caption{Variation of $n=0 ,l=2$ scalar, EM, and gravitational QNMs with $\lambda$ and $\Gamma$. (a,b) $\lambda=1$ and varying $\Gamma$, (c,d) $\lambda=1/2$ and varying $\Gamma$, (e,f) $\lambda=2/3$ and varying $\Gamma$}.
    \label{fig:three}
\end{figure}

Figure \ref{fig:three} shows the variation of scalar, EM, and gravitational QNMs with $\lambda$ and $\Gamma$. With $\lambda = 1$, Re$(\omega)$ increases with increasing $\Gamma$ for all types of perturbations. However, the change is more dynamic in the case of Im$(\omega)$. The values decrease up to approximately $\Gamma = 0.7$, following which an increasing trend can be observed, and the values tend to converge as $\Gamma \rightarrow 1$. Further, gravitational perturbations are retained longer, as Im$(\omega)_{grav}>$ Im$(\omega)_{EM}$ and Im$(\omega)_{saclar}$. Next, with $\lambda = 1/2$, we observe similar trends for Re$(\omega)$; however, in this case, the difference in Re$(\omega)_{grav}$ is much more prominent, whereas that for the scalar and EM cases are nearly identical. For  Im$(\omega)$, we have an increasing behavior with increasing $\Gamma$, and the difference in  Im$(\omega)_{grav}$ is more prominent than that for the scalar and EM cases, which are nearly identical. Further, the gravitational mode in this case is the slowest decaying among all the sets of parameters. Further, the qualitative change the QNM frequencies with changing $\Gamma$ with $\lambda = 2/3$ are approximately identical. For all perturbations, higher values of $\Gamma$ tend to exhibit noticeable deviations from the Schwarzschild case when $\lambda = 1$. However, for the  $\lambda = 1/2$ and  $\lambda = 2/3$ cases, differences from the Schwarzschild case are prominent even with lower values of $\Gamma$.
\begin{table}[!h]
    \begin{subtable}[t]{0.45\textwidth}
    \centering
         \adjustbox{width=\textwidth}{
         \begin{tabular}[t]{|ccc|}

        \hline
        \text{n} & $\omega$  & \text{error estimation} \\
        \hline
        0 & $0.49205\, -0.0972679 i$ & $3.02785\times 10^{-7}$ \\
        1 & $0.472656\, -0.297019 i$ & $0.0000103502$ \\
        2 & $0.440126\, -0.510594 i$ & $0.000492148$ \\
        3 & $0.405992\, -0.739509 i$ & $0.00759531$ \\
        4 & $0.380145\, -0.971383 i$ & $0.047772$ \\
        5 & $0.359138\, -1.19336 i$ & $0.1426$ \\
        6 & $0.349324\, -1.41132 i$ & $0.266999$ \\
        7 & $0.362778\, -1.62911 i$ & $0.400426$ \\
        8 & $0.404289\, -1.84992 i$ & $0.546448$ \\
        9 & $0.448448\, -1.81085 i$ & $1.10486$ \\
        10 & $0.563954\, -2.05955 i$ & $1.29856$ \\
        \hline
   
             \end{tabular}
             }
              \caption{Scalar QNMs at varying overtones}
    \end{subtable}
\hspace{\fill}
    \begin{subtable}[t]{0.45\textwidth}
        \centering
         \adjustbox{width=\textwidth}{
\begin{tabular}[t]{|ccc|}
\hline
 \text{n} & $\omega$  & \text{error estimation} \\
 \hline
 0 & $0.465868\, -0.0955475 i$ & $4.2845\times10^{-7}$ \\
 1 & $0.445234\, -0.292222 i$ & 0.0000141037 \\
 2 & $0.410697\, -0.503765 i$ & 0.000646511 \\
 3 & $0.375052\, -0.73177 i$ & 0.00958649 \\
 4 & $0.34927\, -0.962523 i$ & 0.0559443 \\
 5 & $0.32952\, -1.18486 i$ & 0.153359 \\
 6 & $0.322531\, -1.40601 i$ & 0.274343 \\
 7 & $0.29225\, -1.41738 i$ & 0.738059 \\
 8 & $0.34185\, -1.6422 i$ & 0.879795 \\
 9 & $0.421087\, -1.88112 i$ & 1.06579 \\
 10 & $0.533435\, -2.14096 i$ & 1.31861 \\
 \hline
\end{tabular}
}

    \caption{EM QNMs for varying overtones using Pad\'{e} averaged WKB approach with $l=2$}
    \label{tab:newfour}
     \end{subtable}
     \begin{subtable}[t]{0.45\textwidth}
    \centering
         \adjustbox{width=\textwidth}{
         \begin{tabular}[t]{|ccc|}

        \hline
        \text{n} & $\omega$  & \text{error estimation} \\
        \hline
        0 & $0.437716\, -0.0935634 i$ & $7.19913\times10^{-7}$ \\
        1 & $0.415621\, -0.28668 i$ & 0.0000124227 \\
        2 & $0.378694\, -0.495858 i$ & 0.000880327 \\
        3 & $0.341268\, -0.722432 i$ & 0.012862 \\
        4 & $0.314527\, -0.94971 i$ & 0.0703929 \\
        5 & $0.293838\, -1.16965 i$ & 0.176377 \\
        6 & $0.240418\, -1.19631 i$ & 0.617991 \\
        7 & $0.268203\, -1.41612 i$ & 0.743874 \\
        8 & $0.322646\, -1.64798 i$ & 0.906756 \\
        9 & $0.408134\, -1.89967 i$ & 1.13302 \\
        10 & $0.528267\, -2.17817 i$ & 1.44798 \\
        \hline
   
             \end{tabular}
             }
              \caption{Grav QNMs at varying overtones}
    \end{subtable}

     \caption{QNMs with the Pad\'{e}-averaged WKB method with varying overtones for the modified black hole.}
     \label{tab:newthree}
\end{table}

\subsection{QNMs for the GUP--corrected KR black hole}
\label{sec:gupqnm}
\begin{figure}[!h]
    \centering
    \includegraphics[width=0.9\textwidth]{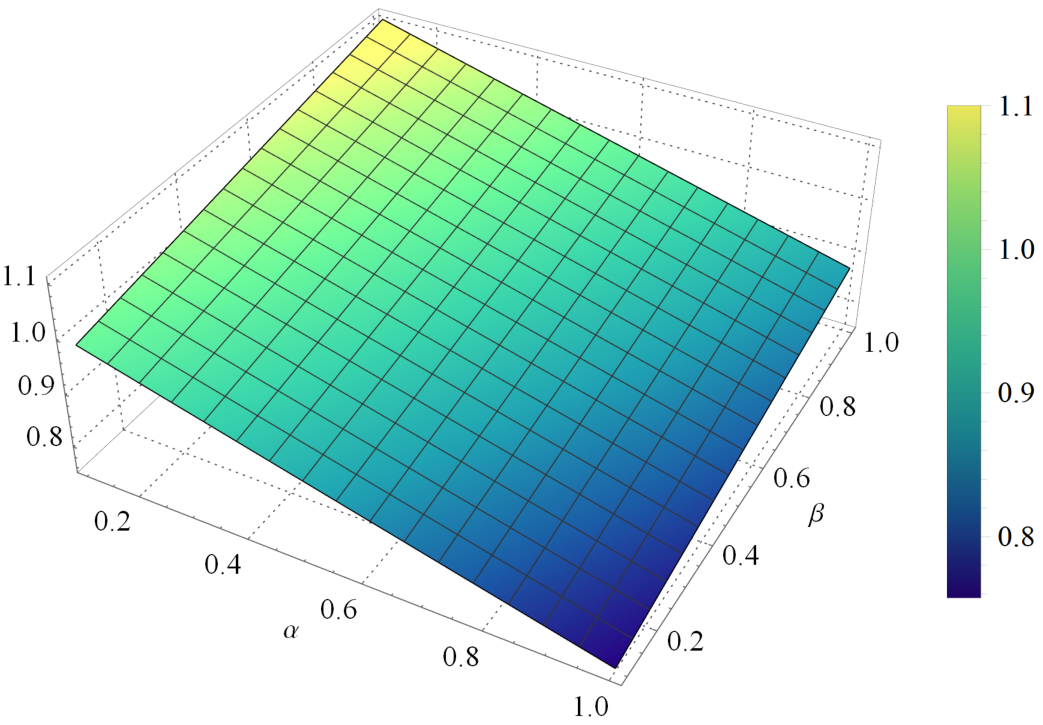}
    \caption{\textit{M\textsubscript{GUP}} for a range of $\alpha$ and $\beta$. Here, we set the LV parameters $\Gamma$ and $\lambda$ to 0.1 and 1, respectively.}
    \label{fig:four}
\end{figure}

Here, we focus on the QNMs evaluated for the GUP--corrected black hole space--time given by the metric Eq. \eqref{GUPmetric}. First, our primary goal here is to check for variations with respect to the LV parameters, and we set $\alpha=\beta=0.1$. We show the changes in the GUP--corrected mass for arbitrary values of $\alpha$ and $\beta$ in Fig. \ref{fig:four} for clarity.

Tables \ref{tab:GUPQNM} \textbf{(a), (b)} and \textbf{(c)} show the $n=0, l=2$ QNMs for scalar, EM, and gravitational perturbations of the GUP--corrected black hole, respectively. It can be seen that both Re$(\omega)$ and Im$(\omega)$ are slightly higher than that those for the modified black hole in Eq. \eqref{eq:BHle} for all types of perturbations. As in the case described before, gravitational perturbations are retained longer, and the GUP--corrected black hole retains perturbations slightly longer than that for the modified black hole in Eq. \eqref{eq:BHle}. The data again are approximately identical to those with the $\Gamma=\lambda=0$ (Schwarzschild) case, with only slight variations. We also show the estimated QNMs at different multipole numbers in subtable \textbf{(d)}, where we have good bounds on the error with increasing $l$. Additionally, we show the QNMs estimated at varying overtones in Table \ref{tab:mainthree}, where these trends are also confirmed. Next, we wish to check the variations of the QNMs with respect to the LV parameters for perturbations of the GUP--corrected black hole.

\begin{table}[htbp]
    \begin{subtable}[t]{0.45\textwidth}
        \centering
             \adjustbox{width=\textwidth}{\begin{tabular}{|ccc|}

\hline
 \text{order} & \text{frequency} & \text{error estimation} \\
 \hline
 1 & $0.503532 - 0.0945403 i$ & 0.0180605 \\
 2 & $0.498507 - 0.0980382 i$ & 0.00117126 \\
 3 & $0.498428 - 0.0984571 i$ & 0.0000192727 \\
 4 & $0.498504\, -0.0985054 i$ & $9.55612\times10^{-6}$ \\
 5 & $0.498504\, -0.0985138 i$ & $4.27083\times10^{-6}$ \\
 6 & $0.498499\, -0.0985112 i$ & $5.83701\times10^{-7}$ \\
 7 & $0.4985\, -0.0985108 i$ & $9.02315\times10^{-8}$ \\
 8 & $0.498498\, -0.0985108 i$ & $2.21541\times10^{-6}$ \\
 9 & $0.4985\, -0.0985108 i$ & $7.76139\times10^{-8}$ \\
 10 & $0.498499\, -0.0985108 i$ & $1.44843\times10^{-6}$ \\
 11 & $0.4985\, -0.0985109 i$ & $2.6724\times10^{-7}$ \\
 12 & $0.498499\, -0.0985123 i$ & $2.66948\times10^{-6}$ \\
 13 & $0.4985\, -0.0985108 i$ & $3.11896\times10^{-7}$ \\
\hline
\end{tabular}

    }
      \caption{\small{Variation of $l=2, n=0$ scalar QNMs with the Pad\'{e}-averaged WKB method for the GUP--corrected black hole}}
    \end{subtable}
    \begin{subtable}[t]{0.45\textwidth}
        \centering
         \adjustbox{width=\textwidth}{
      \begin{tabular}{|ccc|}

\hline
 \text{order} & \text{frequency} & \text{error estimation} \\
 \hline

 1 & $0.477283\, -0.0926288 i$ & 0.0183124 \\
 2 & $0.472014\, -0.0962464 i$ & 0.00121988 \\
 3 & $0.47191\, -0.0967119 i$ & 0.0000189638 \\
 4 & $0.471987\, -0.0967639 i$ & $9.89135\times10^{-6}$ \\
 5 & $0.471986\, -0.0967726 i$ & $4.09322\times10^{-6}$ \\
 6 & $0.471982\, -0.0967698 i$ & $5.99833\times10^{-7}$\\
 7 & $0.471982\, -0.0967695 i$ & $9.30835\times10^{-8}$ \\
 8 & $0.471982\, -0.0967699 i$ & $5.99222\times10^{-7}$ \\
 9 & $0.471984\, -0.0967724 i$ & $2.69116\times10^{-6}$ \\
 10 & $0.471982\, -0.096771 i$ & $2.17995\times10^{-6}$ \\
 11 & $0.471984\, -0.0967734 i$ & $3.05041\times10^{-6}$ \\
 12 & $0.471982\, -0.096771 i$ & $2.14784\times10^{-6}$ \\
 13 & $0.471982\, -0.0967696 i$ & $4.33892\times10^{-7}$ \\

\hline
\end{tabular}
    }
    \caption{\small{Variation of EM QNMs with Pad\'{e} averaged WKB order with $l=2, n=0$ for the GUP--corrected black hole}}
     \end{subtable}

     \begin{subtable}[t]{0.45\textwidth}
            \centering
         \adjustbox{width=\textwidth}{

    \begin{tabular}{|ccc|}

\hline
 \text{order} & \text{frequency} & \text{error estimation} \\
 \hline
 1 & $0.449058\, -0.0904231 i$ & 0.0185732 \\
 2 & $0.443513\, -0.0941735 i$ & 0.00127575 \\
 3 & $0.443372\, -0.0947003 i$ & 0.0000181346 \\
 4 & $0.443448\, -0.0947542 i$ & \text{$9.55091\times10^{-6}$} \\
 5 & $0.443444\, -0.0947598 i$ & $2.4117\times10^{-7}$\\
 6 & $0.443444\, -0.0947594 i$ & $6.4422\times10^{-8}$ \\
 7 & $0.443444\, -0.0947594 i$ & $6.56843\times10{^-9}$ \\
 8 & $0.443444\, -0.0947594 i$ & $6.50456\times10^{-8}$ \\
 9 & $0.443444\, -0.0947593 i$ & $3.61449\times10^{-8}$ \\
 10 & $0.443444\, -0.0947601 i$ & $1.20882\times10^{-6}$ \\
 11 & $0.443444\, -0.0947593 i$ & $2.08791\times10^{-8}$ \\
 12 & $0.443444\, -0.0947602 i$ & $1.38126\times10^{-6}$ \\
 13 & $0.443444\, -0.0947596 i$ & $6.9656\times10^{-7}$ \\

\hline
\end{tabular}
    
    }
    \caption{\small{Gravitational QNMs using the Pad\'{e}-averaged WKB method with $l=2, n=0$ and $\Gamma = 0.1, \lambda =1$ for the GUP--corrected black hole}}
       \end{subtable}
       \begin{subtable}[t]{0.45\textwidth}
            \centering
         \adjustbox{width=\textwidth}{

    \begin{tabular}{|ccc|}
\hline
 \text{$l$} & \text{frequency} & \text{error estimation} \\
 \hline
  \multicolumn{3}{|c|}{Scalar}\\
   \cline{1-3}
    3 & $0.696083\, -0.0982533 i$ & $3.5032\times10^{-8}$ \\
    4 & $0.894008\, -0.0981461 i$ & $6.56455\times10^{-9}$ \\
    5 & $1.09209\, -0.0980916 i$ & $1.6296\times10^{-9}$ \\
    6 & $1.29025\, -0.0980601 i$ & $5.33493\times10^{-10}$ \\
 \hline\hline
  \multicolumn{3}{|c|}{EM}\\
   \hline
    3 & $0.67728\, -0.0973757 i$ & $4.4103\times10^{-8}$ \\
    4 & $0.879427\, -0.0976176 i$ & $7.44806\times10^{-9}$ \\
    5 & $1.08017\, -0.0977386 i$ & $1.83663\times10^{-9}$ \\
    6 & $1.28018\, -0.0978077 i$ & $5.78444\times10^{-9}$ \\
\hline\hline
  \multicolumn{3}{|c|}{Gravitational}\\
   \hline
    3 & $0.657424\, -0.0964151 i$ & $4.60497\times10^{-8}$ \\
    4 & $0.864145\, -0.0970517 i$ & $8.42096\times10^{-9}$ \\
    5 & $1.06774\, -0.0973648 i$ & $1.48944\times10^{-8}$ \\
    6 & $1.26968\, -0.0975421 i$ & $6.30753\times10^{-10}$ \\
 \hline

    \end{tabular}
    
    }
    \caption{\small{$n=0$ QNMs with varying $l$ and $\Gamma = 0.1, \lambda =1$}}
    \label{tab:ss}
       \end{subtable}
     \caption{$n=0$ QNMs with the Pad\'{e}-averaged WKB method}
     \label{tab:GUPQNM}
\end{table}
\begin{table}[H]
    \begin{subtable}[t]{0.45\textwidth}
    \centering
         \adjustbox{width=\textwidth}{\begin{tabular}{|ccc|}
    \hline
 \text{n} & $\omega$  & \text{error estimation} \\
 \hline
 0 & $0.4985\, -0.0985108 i$ & $3.11896\times10^{-7}$ \\
 1 & $0.47887\, -0.30081 i$ & 0.0000118813 \\
 2 & $0.445936\, -0.517095 i$ & 0.000501567 \\
 3 & $0.411389\, -0.748988 i$ & 0.00756157 \\
 4 & $0.385688\, -0.984527 i$ & 0.0466342 \\
 5 & $0.366213\, -1.21072 i$ & 0.138356 \\
 6 & $0.357737\, -1.43309 i$ & 0.259614 \\
 7 & $0.371725\, -1.65565 i$ & 0.392517 \\
 8 & $0.413493\, -1.88172 i$ & 0.542181 \\
 9 & $0.486537\, -2.11705 i$ & 0.728588 \\
 10 & $0.568864\, -2.09624 i$ & 1.33487 \\
  \hline
    \end{tabular}

             }
    \caption{\small{Scalar QNMs at varying overtones for the GUP--corrected black hole with $l=2$}}
    \end{subtable}
\hspace{\fill}
    \begin{subtable}[t]{0.45\textwidth}
        \centering
         \adjustbox{width=\textwidth}{    \begin{tabular}{|ccc|}
        \hline
        \text{n} & $\omega$  & \text{error estimation} \\
        \hline
        0 & $0.471982\, -0.0967696 i$ & $4.33892\times10^{-7}$ \\
        1 & $0.451101\, -0.295957 i$ & 0.0000117107 \\
        2 & $0.41617\, -0.510196 i$ & 0.000649771 \\
        3 & $0.380126\, -0.741045 i$ & 0.00972451 \\
        4 & $0.354085\, -0.974614 i$ & 0.0567724 \\
        5 & $0.334321\, -1.19956 i$ & 0.155635 \\
        6 & $0.327721\, -1.42309 i$ & 0.278936 \\
        7 & $0.344488\, -1.64862 i$ & 0.416929 \\
        8 & $0.351007\, -1.65984 i$ & 0.907312 \\
        9 & $0.433525\, -1.90634 i$ & 1.1032 \\
        10 & $0.550497\, -2.17549 i$ & 1.37491 \\
        \hline
    \end{tabular}

}
    \caption{\small{EM QNMs at varying overtones for the GUP--corrected black hole $l=2$}}
     \end{subtable}
     \hspace{\fill}
     \centering
\begin{subtable}[t]{0.45\textwidth}
        \centering
         \adjustbox{width=\textwidth}{    \begin{tabular}[t]{|ccc|}
        \hline
         \text{n} & $\omega$  & \text{error estimation} \\
        \hline
        0 & $0.443444\, -0.0947596 i$ & $6.9656\times10^{-7}$ \\
        1 & $0.421083\, -0.290347 i$ & 0.0000112297 \\
        2 & $0.383723\, -0.502183 i$ & 0.000877242 \\
        3 & $0.345869\, -0.731631 i$ & 0.012895 \\
        4 & $0.318902\, -0.961799 i$ & 0.0709022 \\
        5 & $0.298095\, -1.18442 i$ & 0.178103 \\
        6 & $0.293873\, -1.40749 i$ & 0.304515 \\
        7 & $0.314446\, -1.63402 i$ & 0.449489 \\
        8 & $0.362911\, -1.8709 i$ & 0.635452 \\
        9 & $0.415888\, -1.92134 i$ & 1.15764 \\
        10 & $0.538424\, -2.20519 i$ & 1.47887 \\
        \hline
        \end{tabular}

}
    \caption{\small{Gravitational QNMs at varying overtones for the GUP--corrected black hole $l=2$}}
     \end{subtable}
     \caption{QNMs with the Pad\'{e}-averaged WKB method with varying overtones for the GUP--corrected black hole}
     \label{tab:mainthree}
\end{table}
\subsubsection{Variation of QNMs with $\Gamma$ and $\lambda$ for the GUP corrected black hole}
\begin{figure}[htbp]
    \centering
    \includegraphics[width=0.9\textwidth]{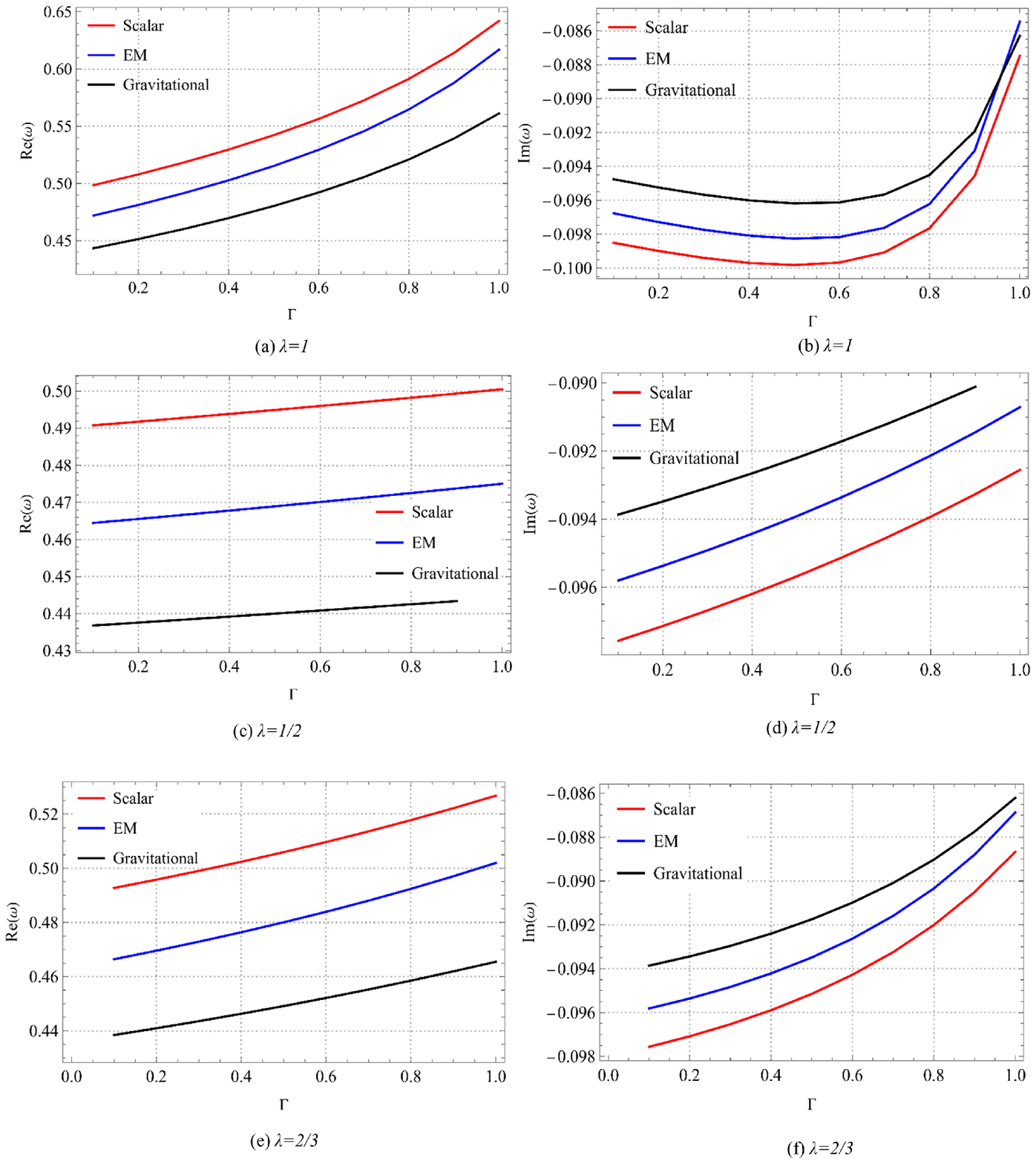}
    \caption{Variation of scalar, EM, and gravitational QNMs with the LV parameters for the GUP--corrected black hole.}
    \label{fig:five}
\end{figure}

In Fig. \ref{fig:five}, we show the variation of the scalar, EM, and gravitational QNMs with the LV parameters for the GUP--corrected black hole. 
With $\lambda = 1$, Re$(\omega)$ for scalar perturbations increases with increasing $\Gamma$; a similar trend can be observed for $\lambda = 1/2$ and $\lambda = 2/3$, but the frequencies for the different perturbations are more closely spaced for the latter two. For Im$(\omega)$ with $\lambda =1$, the values decrease with $\Gamma$ up to approximately $\Gamma = 0.7$, following which we see an increase. For the cases with $\lambda = 1/2$ and $\lambda = 2/3$, the imaginary parts of the QNMs increase with $\Gamma$. Gravitational perturbations exhibit higher values of Im$(\omega)$ for all parametrizations, indicating that they are the slowest decaying among the three. Comparing with the corresponding analysis for the modified black hole shown in Fig. \ref{fig:three}, it can be seen that the frequencies are higher in the GUP--corrected black hole, implying that perturbatins are retained longer in the GUP--corrected black hole.

\subsubsection{Variation of the QNMs with the GUP--parameters $\alpha$ and $\beta$}
\begin{figure}[htbp]
    \centering
\includegraphics[width=0.9\textwidth]{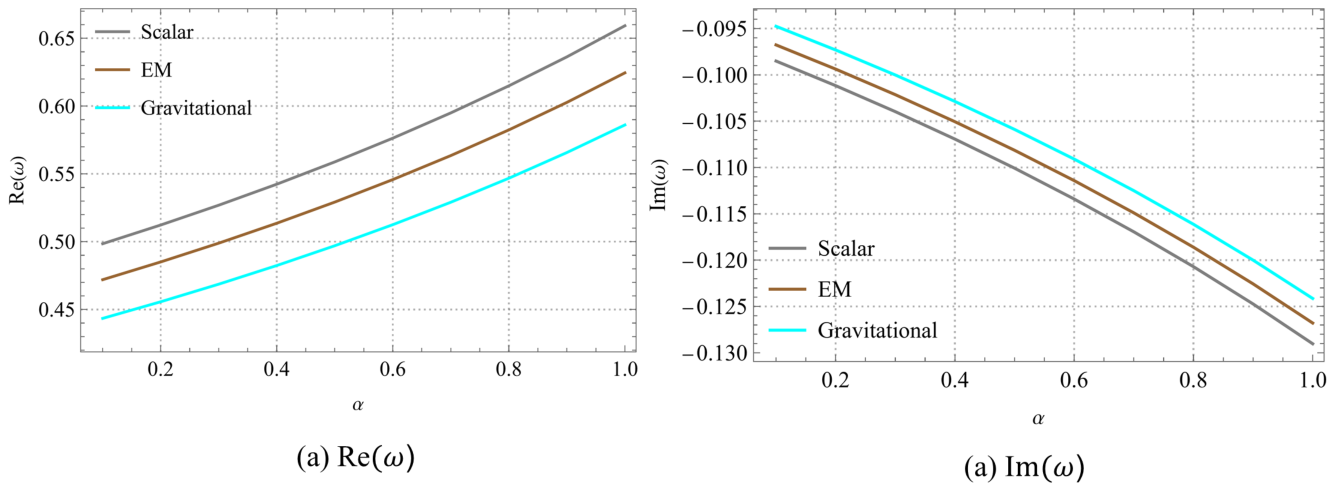}
    \caption{Variation of scalar, EM, and gravitational QNMs the GUP--parameter $\alpha$.}
    \label{fig:six}
\end{figure}

\begin{figure}[htbp]
    \centering
    \includegraphics[width=0.9\textwidth]{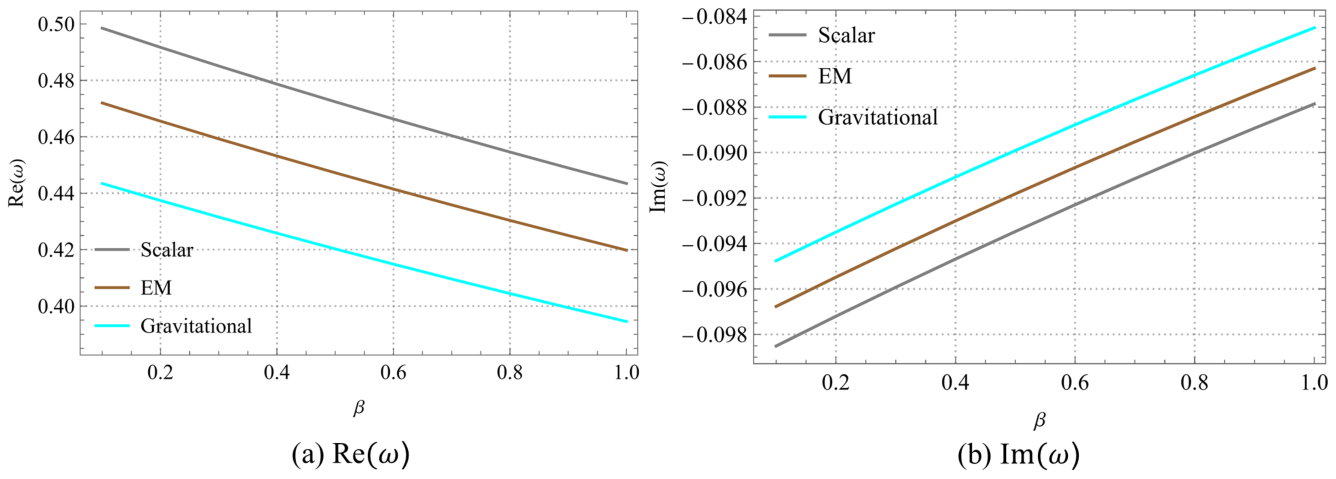}
    \caption{Variation of scalar, EM, and gravitational QNMs the GUP--parameter $\beta$.}
    \label{fig:seven}
\end{figure}

To check the variation of the QNMs with $\alpha$ and $\beta$, we have evaluated the $n=0$ QNMs for $\alpha, \beta \in (0.1,1)$. In Fig. \ref{fig:six}, we show the variation of the QNMs with $\alpha$; here we have considered the $\lambda = 1$ and $\Gamma = 0.1$ case. The frequencies are nearly identical to the case with varying $\Gamma$. Considering the change in Im$(\omega)$, it can be seen that the values for scalar, EM, and gravitational perturbations are nearly identical. However, it is worth noting here that increasing $\alpha$ decreases Im$(\omega)$, implying that the perturbations decay faster with increasing $\alpha$.
Next, in Fig. \ref{fig:seven}, we show the variation of the QNM frequencies with $\beta$ for the three perturbations. It can be seen that Re$(\omega)$ decreases with increasing $\beta$, whereas Im$(\omega)$ increases with increasing $\beta$. Overall, the frequencies are nearly identical with those shown in Fig. \ref{fig:five}, and we have slower decaying perturbations as $\beta$ increases. 

\section{Bounds on the greybody factors}
Black holes are expected to emit thermal radiation outside their event horizons, which is described by the well--known phenomenon of Hawking radiation \cite{hawking1974black, hawking1975particle}. The Hawking spectra as observed by an observer at asymptotic infinity are described by a modified black body distribution, referred to in literature as a greybody distribution. The greybody factors are essential in formulating a general description of the spectrum observed by an observer at infinity \cite{Fernando:2004ay,Boonserm:2019mon,Barman:2019vst,Javed:2022kzf,Okyay:2021nnh,Javed:2021ymu,Al-Badawi:2022aby,Al-Badawi:2021wdm,Mangut:2023oxa}. Here, we focus on estimating the lower bound on the greybody factors of the modified black hole and GUP--corrected black hole. Methods to evaluate the greybody factors include the WKB method and matching technique. A more rigorous analytical bound was proposed initially in Refs. \cite{visser1999some, boonserm2008bounding} as follows:
\begin{equation}
    \mathcal{T} \geq \text{sech}^2 \left(\int_{-\infty}^{\infty} \vartheta dr_{\star} \right)
\end{equation}

Here, $\mathcal{T}$, the transmission probability, is the greybody factor and $\vartheta$ is given by
\begin{equation}
\vartheta = {\sqrt{ (h')^2 + [\omega^2-V- h^2]^2}\over2 h },
\label{eq:varphi}
\end{equation}

where $h(r_{*}) > 0$, satisfying the boundary conditions $h(-\infty)=h(+\infty)=\omega$ is an arbitrary function of the tortoise coordinate defined earlier. In this asymptotic limit $h(r_*) = \omega$, we have
\begin{equation}
\mathcal{T} \geq \text{sech}^2\left( {1\over2\omega} \int_{-\infty}^{\infty} |V(r_*)| \; d r_* \right),
\end{equation}
\begin{equation}
\mathcal{T} \geq \operatorname{sech}^{2}\left(\frac{1}{2 \omega} \int_{r_{H}}^{\infty} \frac{|V|}{f(r)} d r\right),
\end{equation}

Using this, we obtain bounds for $\mathcal{T}_b^{\text{s}}$, $\mathcal{T}_b^{\text{e}}$, and $\mathcal{T}_b^{\text{g}}$, the greybody factors corresponding to the effective potentials for the scalar, EM, and gravitational cases as follows:
\begin{figure}[htbp]
    \centering
    \includegraphics[width=0.9\textwidth]{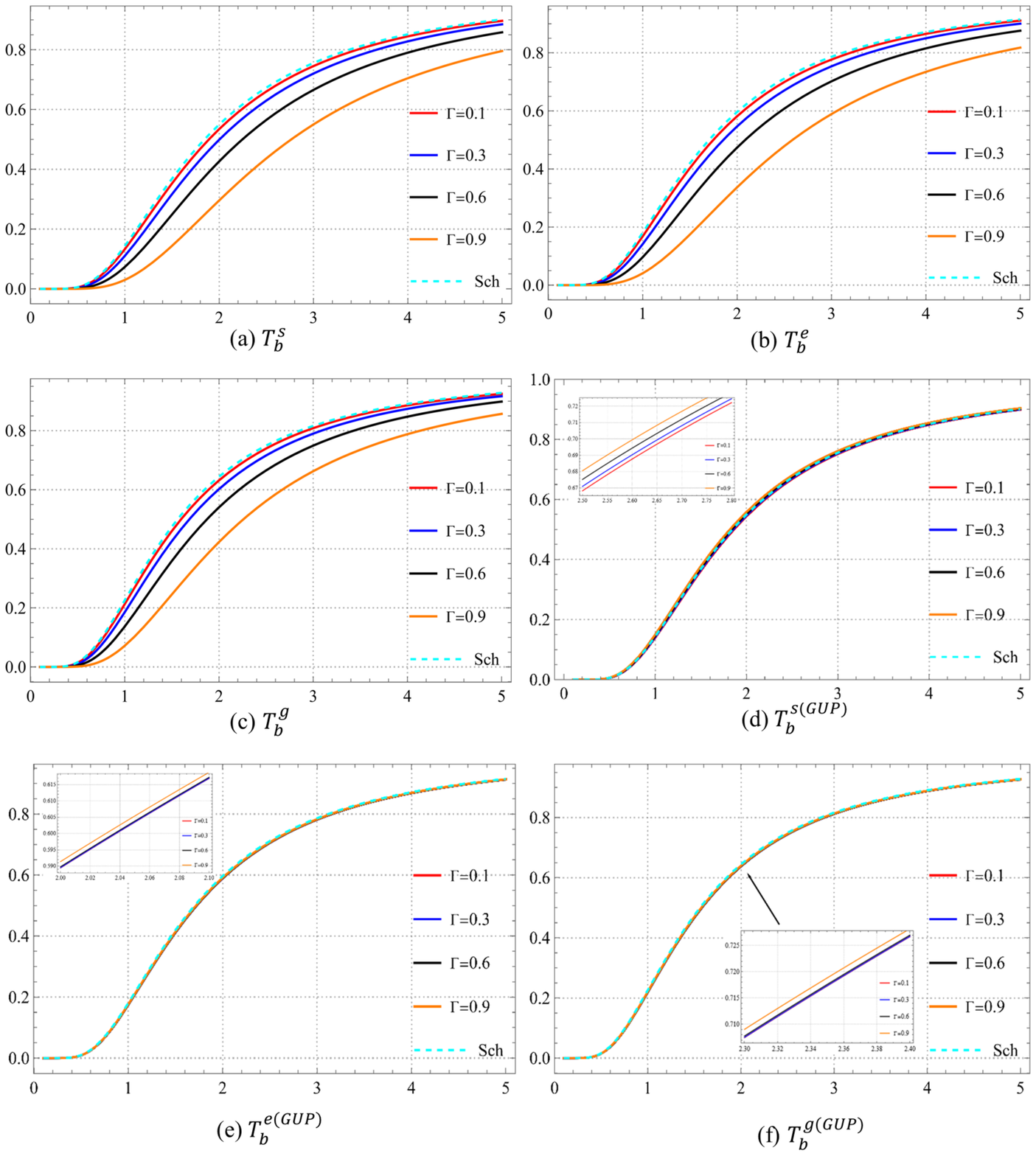}
    \caption{Bounds for the grebody factors of the modified black hole and GUP--corrected black hole. (a) Scalar mode for the modified black hole, (b) EM mode for the modified black hole, (c) Gravitational mode for the modified black hole, (d) Scalar mode for the GUP--corrected black hole, (e) EM mode for the GUP--corrected black hole, (f) Gravitational mode for the GUP--corrected black hole. The horizontal axis denotes $\omega$ and the vertical axis denotes $T$. The insets in (d-f) show the variations in small intervals.}
    \label{fig:gbf}
\end{figure}
\begin{equation}
    \mathcal{T} \geq \mathcal{T}_b^{\text{s}} = \text{sech}^2 \left\{\frac{1}{2\omega} \frac{l^2 \left(\sqrt{M^2-\Gamma }+M\right)+l \left(\sqrt{M^2-\Gamma }+M\right)-\frac{2 \Gamma  \left(\sqrt{M^2-\Gamma }+M\right)^{\frac{\lambda -2}{\lambda }}}{\lambda +2}+M}{\left(\sqrt{M^2-\Gamma }+M\right)^2}\right\},
\end{equation}
which correctly reduces to the Schwarzschild bound $T \geq \text{sech}^2 \left\{\frac{2l(l+l)+1}{8\omega M} \right\}$ as $\lambda \rightarrow 0$, $\Gamma \rightarrow 0$. Similarly, for the EM and gravitational case, we have:
\begin{equation}
\mathcal{T} \geq \mathcal{T}_b^{\text{e}} = \text{sech}^2 \left\{\frac{1}{2\omega} \frac{l (l+1)}{\sqrt{M^2-\Gamma }+M} \right\},
\end{equation}
\begin{align}
    \mathcal{T} \geq \mathcal{T}_b^{\text{g}} = \text{sech}^2 \left \{\frac{1}{2\omega} \frac{M \left(l^2+l+\frac{2 \Gamma  (\lambda -1) \left(\sqrt{M^2-\Gamma }+M\right)^{-2/\lambda }}{\lambda +2}-1\right)}{\left(\sqrt{M^2-\Gamma }+M\right)^2} \right. \\ \nonumber
    +\left.\frac{\sqrt{M^2-\Gamma } \left(l^2+l+\frac{2 \Gamma  (\lambda -1) \left(\sqrt{M^2-\Gamma }+M\right)^{-2/\lambda }}{\lambda +2}\right)}{\left(\sqrt{M^2-\Gamma }+M\right)^2}\right\}
\end{align}
Further, for the GUP--corrected black hole, we evaluate the bounds $\mathcal{T}^{s({gup})}_b$, $\mathcal{T}^{e({gup})}_b$, and $\mathcal{T}^{g({gup})}_b$ for scalar, EM, and gravitational effective potentials, respectively. The expressions are cumbersome to display, so we focus instead on the graphical representation shown in Figure \ref{fig:gbf}. It can be seen that the greybody factors for the modified black hole exhibit deviations from the $\Gamma =0, \lambda =0$ (Schwarzschild) case as $\Gamma$ increases, as expected. It is interesting to note that for the GUP--corrected black hole, the variations with $\Gamma$ are indistinguishable. The insets show the variations in small intervals.

\section{Discussions and Conclusions}
\label{sec:conc}
In this study, we have focused on black hole QNMs and greybody factors for GUP--corrected black holes in non--minimally coupled KR gravity. As discussed in the previous sections, black hole QNMs and greybody factors are crucial in constraining possible modifications of GR. Although important aspects of black holes in KR gravity such as gravitational lensing and particle dynamics have been reported previously \cite{chakraborty2017strong, atamurotov2022particle}, this study presents the first analyses on QNMs and greybody factors of quantum corrected black holes in KR gravity.

In order to incorporate quantum corrections, we have explored the geometric properties of the horizon and QNMs by introducing the concept of the GUP. Comparison of the horizon radii versus the GUP--corrected radii have revealed that the GUP--corrected horizon is smaller in size compared to the horizon of the modified KR black hole. It is worth noting that in contrast to Schwarzschild black holes where GUP--corrections effectively increase the horizon radius, our analyses have demonstrated a reduction in the horizon radius with the GUP corrections. Next, we have leveraged a higher--order Pad\'e--averaged WKB method to estimate the QNMs and used rigorous analytic bounds for the greybody factors. First, we have demonstrated the advantages of the method used here in Section \ref{sec:padeqnm}, where it was observed that the Pad\'e--averaged WKB method exhibits significantly improved bounds on the errors in the estimated frequencies at higher orders, in contrast to the well--known WKB method.

Using this method, we have first evaluated the QNMs for scalar, EM, and gravitational modes for the modified black hole solution in KR gravity without GUP--correction. We have highlighted the effects of the LV parameters on the estimated frequencies, where it was demonstrated that the gravitational perturbations are retained longer for all sets of parametrizations adopted. Next, we have discussed the QNMs for all three perturbations for the GUP--corrected black hole Section \ref{sec:gupqnm}. As for the modified black hole (without GUP--correction), gravitational perturbations have exhibited longer retention, and the GUP--corrected black hole retains perturbations slightly longer than the modified black hole. Here, $Re(\omega)$ exhibits similar dependence on $\Gamma$, but the frequencies for different perturbations are more closely spaced for  $\lambda = 1/2$ and $\lambda = 2/3$. Regarding $Im(\omega)$ with $\lambda = 1$, the values have decreased with increasing $\Gamma$ until approximately $\Gamma = 0.7$, after which an increase has been observed. For the cases of $\lambda = 1/2$ and $\lambda = 2/3$, the imaginary parts of the QNMs have increased with $\Gamma$. Gravitational perturbations are the slowest decaying among the three types in this case as well. Comparing these results with the corresponding analysis for the modified black hole shown in Figure \ref{fig:three}, we have noticed that the frequencies in the GUP--corrected black hole are higher, implying that the GUP--corrected black hole retains perturbations for a longer duration. While the deviations from GR--Schwarzschild data are insignificant for the modified black hole, they are more pronounced for the GUP--corrected solution. However, the differences in the estimated data are not very significant, and consequently, any deviations between the two may prove challenging to verify in future observations.

To examine the variation of the QNMs with $\alpha$ and $\beta$, we have calculated the $n=0$ QNMs for $\alpha, \beta \in (0.1,1)$. It is worth noting that increasing $\alpha$ decreases $Im(\omega)$, indicating faster decaying perturbations. In contrast, $Im(\omega)$ exhibits an increase with higher values of $\beta$. These frequency trends align closely with those depicted in Figure \ref{fig:five}, and we can conclude that perturbations decay at a slower rate as $\beta$ increases. The frequencies remain nearly identical to the previous cases, confirming the consistency of our findings. Additionally, these results establish that the extent of deviations from Schwarzschild behavior are mediated by the GUP--parameters. Interestingly, we have found that the QNM frequencies exhibit a similar dependence on the GUP--parameters $\alpha$ and $\beta$ as those observed in Ref. \cite{gogoi2022quasinormal} for the Bumblebee black hole with topological defects. In accordance with the WKB approximation, it has been observed that the imaginary part of the QNM frequency is consistently negative. This significant observation has indicated the stability of the space--time under perturbations. It has been demonstrated in previous studies that GUP--corrections to black hole space--times are connected with Lorentz violation \cite{tawfik2016lorentz, lambiase2018lorentz}, and our results should serve as an important reference for future studies in this direction. Our findings have also indicated that the gravitational mode is consistently the slowest decaying mode across all cases. This observation highlights the potential for detecting deviations in the ringing behavior of KR black holes in the near future through higher--sensitivity gravitational wave detections.

Moreover, we have estimated the bounds of the greybody factors for scalar, electromagnetic, and gravitational effective potentials in the context of the GUP--corrected black hole. Although the expressions for these bounds are complex, we have provided a graphical representation in Figure \ref{fig:gbf}. The figures illustrate the lower bound of the greybody factor, denoted as $\sigma_l(\omega)$. From the figure, it is evident that $\sigma_l(\omega)$ increases monotonically with the increasing values of $\mathcal{T}^{g({gup})}_b$, while keeping $\Gamma$ fixed. The greybody factors for the modified black hole exhibit deviations from the $\Gamma=0, \lambda=0$ (Schwarzschild) case as $\Gamma$ increases, which aligns with our expectations. Notably, for the GUP-corrected black hole, the variations with $\Gamma$ appear indistinguishable. Thus, our results show the impact of GUP--corrections on the Hawking radiation spectra of LV black hole space--times in KR gravity.

The KR field plays a significant role considering both theoretical and astrophysical motivations. For instance, the KR field can manifest optically active space--times \cite{kroptical}, induce topological defects that can lead to galactic structures with intrinsic angular momentum \cite{krgalactic}, can affect leptogenesis \cite{krlepto} and the cosmic microwave background \cite{krcmb} etc. Moreover, there are strong indications that the KR field can permeate through possible extra dimensions \cite{kred}. Thus, studying the effective field theory of KR gravity is of fundamental importance. Our results show that black holes can serve as a suitable testing ground for probing the existance of the KR field. The results presented in this paper demonstrate the effect of parameters mediating local Lorentz violation and the GUP in KR gravity for the first time. Future analyses, including time--domain calculations and innovative approaches such as spectral methods could provide further insights into the behavior of perturbations in these space--times. Additionally, exploring additional parameters of interest such as the shadows of KR--modified black holes and GUP--corrected black holes within this scenario present new prospects in these space--times. Currently, we are working on these aspects of the research.

\section*{Acknowledgement}

A. {\"O}. would like to acknowledge the contribution of the COST Action
CA18108 - Quantum gravity phenomenology in the multi-messenger approach
(QG-MM).  A. {\"O}. would like to acknowledge the contribution of the COST Action CA21106 - COSMIC WISPers in the Dark Universe: Theory, astrophysics and experiments (CosmicWISPers). A. {\"O}. is funded by the Scientific and Technological Research Council of Turkey (TUBITAK).

\appendix

\section{QNM data for Schwarzschild black hole}
\label{appenA}
The table below shows the QNM data for the Schwarzschild black hole calculated using the Pad\'{e} averaged-WKB method for scalar, EM, and gravitational perturbations at different overtones.
\begin{table}[H]
    \begin{subtable}[t]{0.5\textwidth}
        \centering
             \adjustbox{width=\textwidth}{\begin{tabular}{|ccc|}
\hline
 \text{n} & $\omega$  & \text{error estimation} \\
 \hline
 0 & $0.483644\, -0.0967588 i$ & $3.6171\times 10^{-7}$ \\
 1 & $0.463843\, -0.295611 i$ & 0.000013064 \\
 2 & $0.430649\, -0.50862 i$ & 0.000546428 \\
 3 & $0.395981\, -0.737192 i$ & 0.00848474 \\
 4 & $0.369671\, -0.968109 i$ & 0.0522682 \\
 5 & $0.348501\, -1.18989 i$ & 0.14967 \\
 6 & $0.340115\, -1.40927 i$ & 0.272361 \\
 7 & $0.355156\, -1.62968 i$ & 0.40368 \\
 8 & $0.397612\, -1.85446 i$ & 0.551586 \\
 9 & $0.439328\, -1.82418 i$ & 1.11679 \\
 10 & $0.554535\, -2.08059 i$ & 1.32654 \\
\hline
\end{tabular}}
      \caption{Scalar QNMs for the Schwarzschild black hole}
       \label{tab:scalarSch}
    \end{subtable}
    \hfill
    \begin{subtable}[t]{0.5\textwidth}
        \centering
         \adjustbox{width=\textwidth}{
      \begin{tabular}{|ccc|}
\hline
 \text{n} & $\omega$  & \text{error estimation} \\
 \hline
 0 & $0.457596\, -0.0950046 i$ & $4.9671\times 10^{-7}$ \\
 1 & $0.436529\, -0.290719 i$ & 0.0000142785 \\
 2 & $0.401304\, -0.501664 i$ & 0.000700978 \\
 3 & $0.365144\, -0.72941 i$ & 0.0103468 \\
 4 & $0.339262\, -0.959536 i$ & 0.0593261 \\
 5 & $0.31948\, -1.1814 i$ & 0.158808 \\
 6 & $0.313268\, -1.40253 i$ & 0.28075 \\
 7 & $0.330326\, -1.62599 i$ & 0.417406 \\
 8 & $0.337797\, -1.64074 i$ & 0.898291 \\
 9 & $0.419236\, -1.88605 i$ & 1.09594 \\
 10 & $0.534507\, -2.15433 i$ & 1.37022 \\
 \hline
    \end{tabular}
    }
    \caption{EM QNMs for the Schwarzschild black hole}
    \label{tab:EMSch}  
     \end{subtable}

     \begin{subtable}[t]{0.5\textwidth}
            \centering
         \adjustbox{width=\textwidth}{

    \begin{tabular}{|ccc|}
\hline
 \text{n} & $\omega$  & \text{error estimation} \\
 \hline
 0 & $0.430559\, -0.0930526 i$ & $1.1774\times 10^{-6}$ \\
 1 & $0.408061\, -0.285261 i$ & 0.0000134829 \\
 2 & $0.370456\, -0.493866 i$ & 0.000945885 \\
 3 & $0.332573\, -0.720367 i$ & 0.013354 \\
 4 & $0.30639\, -0.948101 i$ & 0.0708138 \\
 5 & $0.237535\, -0.966349 i$ & 0.535052 \\
 6 & $0.243168\, -1.18854 i$ & 0.628387 \\
 7 & $0.270029\, -1.41363 i$ & 0.750096 \\
 8 & $0.324194\, -1.64988 i$ & 0.918165 \\
 9 & $0.410557\, -1.90675 i$ & 1.15879 \\
 10 & $0.532766\, -2.1919 i$ & 1.49685 \\
 \hline

    \end{tabular}
    
    }
    \caption{Gravitational QNMs for the Schwarzschild black hole}
    \label{tab:GravSch}
       \end{subtable}
       \begin{subtable}[t]{0.5\textwidth}
            \centering
         \adjustbox{width=\textwidth}{

    \begin{tabular}{|ccc|}
\hline
 \text{$l$} & \text{frequency} & \text{error estimation} \\
 \hline
  \multicolumn{3}{|c|}{Scalar}\\
   \cline{1-3}
    3 & $0.675366\, -0.0964996 i$ & $3.9110\times 10^{-8}$ \\
    4 & $0.867416\, -0.0963917 i$ & $7.9415\times^10{-9}$ \\
    5 & $1.05961\, -0.0963368 i$ & $1.9134\times 10^{-9}$ \\
    6 & $1.25189\, -0.0963051 i$ & $6.1835\times 10^{-10}$ \\
 \hline\hline
  \multicolumn{3}{|c|}{EM}\\
   \hline
    3 & $0.656899\, -0.0956162 i$ & $4.9240\times 10^{-8}$ \\
    4 & $0.853095\, -0.0958599 i$ & $8.6866\times 10^{-9}$ \\
    5 & $1.04791\, -0.0959817 i$ & $2.1623\times 10^{-9}$ \\
    6 & $1.2420\, -0.0960512 i$ & $6.7749\times 10^{-10}$ \\
\hline\hline
  \multicolumn{3}{|c|}{Gravitational}\\
   \hline
    3 & $0.638085\, -0.0946835 i$ & $6.0517\times 10^{-8}$ \\
    4 & $0.838615\, -0.0953104 i$ & $1.0467\times 10^{-8}$ \\
    5 & $1.03613\, -0.0956187 i$ & $2.29767\times10^{-9}$ \\
    6 & $1.23205\, -0.0957933 i$ & $7.19463\times10^{-10}$ \\
 \hline

    \end{tabular}
    
    }
    \caption{$n=0$ QNMs with varying $l$ for the Schwarzschild black hole}
    \label{tab:SchQNMsVarL}
       \end{subtable}
     \caption{$n=0$ QNMs with the Pad\'{e}-averaged WKB method for the Schwarzschild black hole}
     \label{tab:mainone}
\end{table}

\bibliography{apssamp}

\end{document}